%% file: arxiv.tex
\documentclass[%
 aip,
jcp,%
 amsmath,amssymb,
 reprint,%
 groupedaddress,
longbibliography
]{revtex4-1}
\usepackage{graphicx}%
\usepackage{dcolumn}%
\usepackage{bm}%
\usepackage{xcolor}
\usepackage{siunitx}
\usepackage{epstopdf}
\usepackage{sidecap}
\sidecaptionvpos{figure}{c}

\newcommand{\Tr}{\mathrm{Tr}}
\newcommand{\tr}{\mathrm{tr}}
\newcommand{\rd}{\mathrm{d}}
\newcommand{\iu}{\mathrm{i}}

\newcommand{\eu}[1]{\mathrm{e}^{#1}}

\newcommand{\pder}[3][]{\frac{\partial^{#1}{#2}}{\partial{#3}^{#1}}}

\newcommand{\half}{\frac{1}{2}}
\usepackage{soul} %
\usepackage{hyperref}
\hypersetup{colorlinks=true,linkcolor=black,citecolor=black,urlcolor=black} %

\usepackage{braket}

\usepackage{dcolumn} %
\newcolumntype{d}[1]{D{.}{.}{#1}} %

\begin{document}

\title{A mapping approach to surface hopping}%

\author{Jonathan R. Mannouch}
\email{jonathan.mannouch@phys.chem.ethz.ch}
\author{Jeremy O. Richardson}%
\email{jeremy.richardson@phys.chem.ethz.ch}
\affiliation{Laboratory of Physical Chemistry, ETH Z\"{u}rich, 8093 Z\"{u}rich, Switzerland}

\date{\today}%

\begin{abstract}
We present a nonadiabatic classical-trajectory approach that offers the best of both worlds between fewest-switches surface hopping (FSSH) and quasiclassical mapping dynamics. This mapping approach to surface hopping (MASH) propagates the nuclei on the active adiabatic potential-energy surface, like in FSSH\@. However, unlike in FSSH, transitions between active surfaces are deterministic and occur when the electronic mapping variables evolve between specified regions of the electronic phase space. This guarantees internal consistency between the active surface and the electronic degrees of freedom throughout the dynamics. MASH is rigorously derivable from exact quantum mechanics, as a limit of the quantum--classical Liouville equation (QCLE), leading to a unique prescription for momentum rescaling and frustrated hops. Hence, a quantum-jump procedure can in principle be used to systematically converge the accuracy of the results to that of the QCLE\@. This jump procedure also provides a rigorous framework for deriving approximate decoherence corrections similar to those proposed for FSSH\@. We apply MASH to simulate the nonadiabatic dynamics in various model systems and show that it consistently produces more accurate results than FSSH, at a comparable computational cost.
\end{abstract}

\maketitle

\section{\label{sec:intro}Introduction}
Classical-trajectory approaches offer a computationally affordable way of simulating coupled electron--nuclear nonadiabatic dynamics in condensed-phase systems.\cite{Stock2005nonadiabatic} Ehrenfest dynamics is one of the simplest methods; it propagates the nuclear degrees of freedom on a mean-field average of the potential-energy surfaces, weighted according to the time-evolved electronic wavefunction.\cite{Tully1998MQC} However, this approach is generally observed to drastically violate detailed balance,\cite{Parandekar2005} and due to its inability to describe wavepacket branching, simulations of chemical reactions form artificial product states. While coupled-trajectory approaches fix these problems in principle by offering a way of systematically improving the accuracy of the results,\cite{Burghardt1999,Shalashilin2011,Curchod2018} they also forfeit the computational efficiency and simplicity of independent trajectories. There are, however, two established approaches for going beyond Ehrenfest using independent trajectories, commonly referred to as surface hopping and quasiclassical mapping, which differ in how the electronic dependence of the nuclear force is described.

Fewest-switches surface hopping (FSSH)\cite{Tully1990hopping} evolves the nuclei on a single adiabatic surface at any given time and nonadiabatic transitions are captured by stochastic hops between active surfaces with probabilities designed to be consistent with the underlying time-evolved electronic degrees of freedom. By construction, surface-hopping approaches describe the branching of a nuclear wavepacket on passing through a region of nonadiabatic coupling, which is one of the most important features necessary to describe excited-state dynamics in photochemistry. It is also generally observed that this method approximately obeys detailed balance, at least in certain limits.\cite{Schmidt2008}
For these reasons it has become one of the most popular methods for performing nonadiabatic dynamics simulations.\cite{Granucci2001,Barbatti2011,Landry2014,Nelson2014,Barbatti2014newtonX,Goyal2016,Spencer2016,Mai2018SHARC}

Despite the many advantageous features, FSSH is nevertheless known to suffer from a number of significant problems. First, there is the philosophical question of why it is necessary to use stochastic hops to simulate the dynamics, when the time-dependent Schr\"odinger equation is deterministic. Second, FSSH is known to incorrectly capture electronic decoherence that occurs as a result of wavepacket branching.\cite{Tully1990hopping,Bittner1995,Subotnik2016review,Kapral2016} This is known to lead to errors in the population transfer when trajectories pass through multiple coupled regions, resulting in an inability to describe charge transport\cite{Carof19} and the $\Delta^{2}$ dependence of Marcus-theory rates,\cite{Landry2011hopping} where $\Delta$ is the electronic diabatic coupling. In order to fix this, a huge range of decoherence corrections have been suggested.\cite{Hammes-Schiffer1994,Bittner1995,Jasper2005,Granucci2010,Subotnik2011AFSSH,Shenvi2011,Subotnik2011,Jaeger2012,Vindel2021} However, these decoherence corrections are not always guaranteed to improve the accuracy of the obtained results, because none of them has a rigorous derivation. For an extreme example, if a decoherence correction were applied at every time-step, then the electronic wavefunction would continuously collapse back to its initial state, in accordance with the quantum Zeno effect, and FSSH would incorrectly predict that the electronic subsystem was stationary.  

Finally, so-called frustrated hops can arise.  These occur when there is insufficient nuclear kinetic energy for a trajectory to reach a higher energy surface, requiring these attempted hops to be aborted in order to conserve energy. It has been noted that aborting hops leads to an inconsistency within FSSH, because the rate of stochastic transitions between surfaces can no longer match the dynamics of the underlying electronic degrees of freedom.\cite{Fang1999} Some have suggested that frustrated hops should be removed from FSSH by either turning off the coupling for transitions that would give rise to frustrated hops\cite{Fang1999} or allowing a trajectory to hop at later times by invoking the time-energy uncertainty principle.\cite{Jasper2002}
The standard procedure is to allow for frustrated hops and to reverse the momentum when they occur.\cite{Hammes-Schiffer1994}
However, others observe an improvement in the accuracy of the nuclear dynamics when frustrated hops are included but the momentum reversal is not applied at all\cite{Muller1997} or only in certain situations.\cite{Jasper2003,Jain2015hopping2}
Still others note that including momentum reversals improves detailed balance,\cite{Sifain2016} internal consistency\cite{Carof17} %
and reaction rates in the Marcus-theory limit (using decoherence corrections).\cite{Jain2015hopping2} Hence a bigger issue is a general lack of consensus over the correct way of fixing these problems within FSSH and even which aspects of the method are important to retain when doing so.

In order to know precisely how to improve upon FSSH, a rigorous derivation is needed. When trying to derive classical-trajectory approaches, one may start from the quantum--classical Liouville equation (QCLE),\cite{Kapral1999,Shi2004QCLE,Bonella2010} which formally describes the dynamics of quantum electrons coupled to classical nuclei,
but which cannot be solved directly
using independent trajectories.\cite{Kapral1999QCLE,Kelly2012mapping}
Recently a pseudo-derivation of FSSH was obtained from the QCLE,\cite{Subotnik2013} but it required a series of major and minor assumptions that meant several important features of the method, such as frustrated hops and decoherence corrections, could not be derived. A related study demonstrated that FSSH could not be derived from the QCLE without neglecting certain terms.\cite{Kapral2016}
Other surface-hopping approaches have been rigorously derived from the QCLE\cite{Kapral1999,Martens2016} %
or the exact-factorization approach\cite{Pieroni2021}
without the need for such assumptions.
However, these methods tend to be computationally expensive or involve coupled trajectories, %
and thus further approximations were required to obtain more practical independent-trajectory methods.\cite{Ha2018,Martens2019,Martens2020}

An alternative way of going beyond Ehrenfest using independent trajectories is with quasiclassical mapping dynamics. Within these approaches, the electronic subsystem is mapped onto a fictitious system containing continuous degrees of freedom, allowing a classical dynamics to be obtained in which the electronic and nuclear degrees of freedom are treated on an equal-footing.\cite{Stock2005nonadiabatic}
The most common is 
the Meyer--Miller--Stock--Thoss (MMST) mapping, based on a collection of harmonic oscillators.\cite{Meyer1979nonadiabatic,Stock1997mapping}
The symmetrical quasiclassical (SQC) windowing approach\cite{Miller2016Faraday} is built on MMST trajectories, but introduces a discrete representation of the population observables obtained from separating the mapping space into
distinct regions that each correspond to a different state.
Several other mappings have also been proposed,\cite{Thoss1999mapping,Cotton2015spin,Meyer1979spinmatrix,Liu2016,*xin2019,Miller1986fermions,*Li2012fermions,*Sun2021,Lang2021GDTWA}
including a new form of `spin mapping',\cite{spinmap,multispin,JohanPhD} which for two-level systems maps to a spin-$\tfrac{1}{2}$ particle and typically exhibits superior accuracy over other mapping approaches.\cite{spinPLDM1,spinPLDM2,SPMD,nonlinear,FMOclassical,ultrafast,GQME} This improved accuracy of the mapping is in part due to the fact that the identity operator is correctly represented by the number one\cite{identity,FMO,linearized} and that a rigorously obtained optimum value for the so-called zero-point energy parameter is used.\cite{spinmap,multispin,Cotton2013mapping,ultrafast} The size of the mapping space is also the same as that of the electronic subsystem, reducing the possibility of classical trajectories leaking into unphysical regions of the mapping space that have no correspondence with the original subsystem.\cite{Mueller1998mapping}
Although mapping methods generally produce better results than Ehrenfest, they still employ mean-field forces, which cannot capture wavepacket branching.\cite{Miller2009mapping,SPMD}
Additionally, in many cases, there are regions in the mapping space that give rise to negative populations, where the nuclei evolve on inverted potentials.\cite{Bonella2001mapping1,Coronado2001,Bellonzi2016,Cotton2019SQC}
These inverted potentials can become unbounded and lead to trajectories unphysically accelerating off to infinity.

Like many other mapping-based approaches, spin mapping can be rigorously derived, so that the accuracy of the obtained dynamics can, in principle, be systematically improved to that of the QCLE by resampling the mapping variables at intermediate times through so-called quantum-jump procedures.\cite{Hsieh2013FBTS,Huo2012PLDM,spinPLDM2} Performing jumps is, however, computationally expensive and the advantage of developing new and improved mapping approaches is that they will typically require fewer jumps to give accurate results, or even give reasonable predictions without any jumps at all.

The following question therefore arises: is there a best-of-both-worlds approach which combines the desirable properties of surface-hopping dynamics with the rigour of mapping? In this paper, we show that a rigorous mapping approach to surface hopping (MASH) is indeed possible to obtain. The MASH approach is based on spin mapping but windows the spin-sphere into distinct regions, each corresponding to an adiabatic state. This goes beyond the windowing of SQC, in that not only the observables but also the forces are determined by the window currently occupied by the trajectory.
This gives rise to surface-hopping dynamics, where hops between active surfaces are deterministic and occur when the mapping variables evolve between windowed regions, which in contrast to FSSH,\cite{Fang1999} guarantees that there is always internal consistency between the electronic degrees of freedom and the active surface. The approach is rigorously derivable as a short-time approximation to the QCLE and therefore leads to specific and unique prescriptions for how the momentum rescaling and frustrated hops should be performed. A quantum-jump procedure can also be rigorously obtained, allowing systematic convergence of the results to those of the QCLE, at least in principle. This %
provides a formalism from which decoherence corrections, similar to those used in FSSH, can be derived.

In this paper, we start by giving a brief review of FSSH and spin mapping, before outlining our new MASH approach. Finally, we benchmark MASH against other classical-trajectory methods by calculating dynamical observables for a range of commonly used nonadiabatic models.  We find that MASH alleviates many of the known problems of FSSH and therefore consistently produces more accurate results.    

\section{Background Theory}
The nonadiabatic Hamiltonian for a general electron--nuclear coupled system can be written in the form
\begin{equation}
\label{eq:ham}
\hat{H}=\sum_{j=1}^{f}\frac{p_{j}^{2}}{2m}+\bar{V}(\bm{q})+\hat{V}(\bm{q}) ,
\end{equation}
where $\bm{q}=\{q_{1},\dots,q_{f}\}$ and $\bm{p}=\{p_{1},\dots,p_{f}\}$ are the coordinates and momenta associated with the nuclear modes $j$, which are mass-weighted so that all $f$ degrees of freedom have the same mass $m$. $\bar{V}(\bm{q})$ is the state-independent scalar potential, whereas $\hat{V}(\bm{q})$ is the state-dependent potential operator, defined so that $\tr[\hat{V}(\bm{q})]=0$, where $\tr[\cdot]$ is the partial trace over the electronic degrees of freedom. Strictly speaking, Eq.~(\ref{eq:ham}) can only be called the Hamiltonian of the system (such that it generates the dynamics) if the state-dependent potential is expressed in a diabatic electronic basis which is independent of the nuclear phase-space variables.

However, it is often advantageous to work with adiabatic states, so that the electronic states diagonalize the matrix $\hat{V}(\bm{q})$. For two-level problems, which we consider in this paper, the adiabatic electronic states are $\ket{\psi_\pm(\bm{q})}$, in terms of which the state-dependent potential is
\begin{equation}
\label{eq:V_ad}
\hat{V}(\bm{q})= V_{z}(\bm{q})\Big[\ket{\psi_{+}(\bm{q})}\!\bra{\psi_{+}(\bm{q})}-\ket{\psi_{-}(\bm{q})}\!\bra{\psi_{-}(\bm{q})}\Big] ,
\end{equation}
and the corresponding adiabatic potential-energy surfaces are $V_{\pm}(\bm{q})=\bar{V}(\bm{q})\pm V_{z}(\bm{q})$.
We choose $V_z(\bm{q})\ge0$ such that $V_+(\bm{q})$ describes the upper (excited) and $V_-(\bm{q})$ the lower (ground) state. Because the adiabatic states depend explicitly on $\bm{q}$, this means that in this basis the representation of the Hamiltonian in Eq.~(\ref{eq:ham}) no longer generates the dynamics, but still corresponds to a conserved energy. 

One could instead use the full adiabatic representation of the Hamiltonian\cite{Cederbaum2004CI}
\begin{equation}
\label{eq:ham_ad}
\hat{H}^{\text{ad}}=\sum_{j=1}^{f}\frac{\left[\hat{p}^{\text{ad}}_{j}(\bm{q})+d_{j}(\bm{q})\hat{\sigma}_{y}(\bm{q})\right]^{2}}{2m}+\bar{V}(\bm{q})+V_{z}(\bm{q})\hat{\sigma}_{z}(\bm{q}) ,
\end{equation}
where $\hat{p}^{\text{ad}}_{j}(\bm{q})=p_j-d_j(\bm{q})\hat{\sigma}_y(\bm{q})$ is the canonical momentum in the adiabatic representation 
and we set $\hbar=1$ throughout. 
The Pauli spin matrices in the adiabatic basis, $\hat{\bm{\sigma}}(\bm{q})=\{\hat{\sigma}_{x}(\bm{q}),\hat{\sigma}_{y}(\bm{q}),\hat{\sigma}_{z}(\bm{q})\}$, are defined in Appendix~\ref{sec:transf} and along with the identity, $\hat{\mathcal{I}}$, form a complete basis for Hermitian operators corresponding to two-level problems. Additionally, $d_{j}(\bm{q})$ is an element of the nonadiabatic coupling vector, defined as
\begin{equation}
\label{eq:nonadiabatic_coup} 
d_{j}(\bm{q})=\Braket{\psi_{+}(\bm{q})|\frac{\partial\psi_{-}(\bm{q})}{\partial q_{j}}}=-\Braket{\psi_{-}(\bm{q})|\frac{\partial\psi_{+}(\bm{q})}{\partial q_{j}}} .
\end{equation}
This approach is, however, not desirable, because the dynamical equations of motion generated from Eq.~(\ref{eq:ham_ad}) contain the second-derivative couplings, $\braket{\psi_{\pm}(\bm{q})|\partial^{2}\psi_{\pm}(\bm{q})/\partial q_{j}^{2}}$, which are typically difficult to compute. To avoid these complications, %
we choose to work in the kinematic picture\cite{Cotton2017mapping} [Eqs.~(\ref{eq:ham}) and (\ref{eq:V_ad})], which gives rise to identical dynamics as those generated by Eq.~(\ref{eq:ham_ad}), but instead expresses the equations of motion in terms of $p_{j}$, so that they do not explicitly depend on the second-derivative couplings. As an extra advantage, $p_{j}$ retains the intuitive meaning of mass times velocity in this picture. 
This kinematic picture is also implicitly used in the standard surface-hopping approach (as is obvious from the fact that they all employ $\dot{q}_j=p_j/m$).

Most relevant dynamical observables can be written in terms of real-time correlation functions. In this paper, we will consider single-time correlation functions of the form
\begin{equation}
\label{eq:corr_function}
C_{AB}(t)=\Tr\Big[\hat{A}\hat{B}(t)\Big],
\end{equation}
where $\Tr[\cdot]$ is the trace over both the nuclear and electronic degrees of freedom and $\hat{B}(t)$ is a time-evolved quantum operator in the Heisenberg picture. %
This correlation function can be interpreted as measuring the response, $\hat{B}$, at time $t$ after the system is initialized by the operator $\hat{A}$. In general, both $\hat{A}$ and $\hat{B}$ contain nuclear and electronic components.

When considering approximations to Eq.~(\ref{eq:corr_function}) that can be easily computed using independent classical trajectories, it is most natural to represent the operators as functions of the associated nuclear and electronic phase-space variables
\begin{subequations}
\label{eq:corr_phase}
\begin{align}
C_{AB}(t)&\approx\Big\langle A(\bm{q},\bm{p},\bm{S})B(\bm{q}(t),\bm{p}(t),\bm{S}(t))\Big\rangle , \\
\Big\langle\cdots\Big\rangle&=\int\frac{\rd\bm{q}\,\rd\bm{p}}{(2\pi)^{f}}\int\rd\bm{S}\cdots , \label{eq:av_def}
\end{align}
\end{subequations}
where the $\{\bm{q},\bm{p}\}$ dependence of the observable operator functions is generated by the Wigner transform of the nuclear part of the associated operator. In addition, a convenient way of completely describing any pure state of a two-level electronic subsystem is via the electronic phase-space variables, $\bm{S}=\{S_{x},S_{y},S_{z}\}$, defined by
\begin{subequations}
\label{eq:density_elem}
\begin{align}
S_{x}&=2\text{Re}\!\left[c^{*}_{+}c_{-}\right] , \\
S_{y}&=2\text{Im}\!\left[c^{*}_{+}c_{-}\right] , \\
S_{z}&=|c_{+}|^{2}-|c_{-}|^{2} ,
\end{align}
\end{subequations}
which correspond to the expectation values of the Pauli spin matrices with respect to the electronic wavefunction, $\ket{\Psi}=c_{+}\ket{\psi_{+}(\bm{q})}+c_{-}\ket{\psi_{-}(\bm{q})}$. For a normalized electronic wavefunction, $\bm{S}$ lies on the surface of the Bloch sphere with radius one, because $\sqrt{S_{x}^{2}+S_{y}^{2}+S_{z}^{2}}=|c_{+}|^{2}+|c_{-}|^{2}=1$. The form of the operator functions, $A(\bm{S})$, and the electronic phase-space integration measure, $\rd\bm{S}$, are specific to the method and will be defined later when each classical-trajectory approach is introduced. Additionally, the time-evolved operator, $\hat{B}(t)$, is described by a linearized semiclassical (LSC) approximation,\cite{Sun1998mapping,Miller2001SCIVR,Miller2009mapping} $B(\bm{q}(t),\bm{p}(t),\bm{S}(t))$, where the associated electronic and nuclear phase-space variables are evolved by classical dynamics. 
The equations of motion for these dynamics will be presented in the following.

\subsection{The Classical Path Approximation}
We first consider a simpler problem of the bare electronic dynamics along a pre-specified nuclear path [$q_{j}(t)$ with $p_{j}(t)=m\dot{q}_{j}(t)$], which corresponds to the so-called classical path approximation.\cite{Mott1931,Stock2005nonadiabatic} The equations of motion for the electronic phase-space variables are chosen to reproduce the time-dependent Schr{\"o}dinger equation, $\rd\ket{\Psi}/\rd t=-\iu\hat{V}(\bm{q})\ket{\Psi}$, along this path
\begin{subequations}
\label{eq:mapping_eom}
\begin{align}
\dot{S}_{x}&=\frac{2\sum_{j}d_{j}(\bm{q})p_{j}}{m}S_{z}-2V_{z}(\bm{q})S_{y} , \\
\dot{S}_{y}&=2V_{z}(\bm{q})S_{x} , \\
\dot{S}_{z}&=-\frac{2\sum_{j}d_{j}(\bm{q})p_{j}}{m}S_{x} , \label{eq:mapping_eom_z} 
\end{align}
\end{subequations}
and are therefore guaranteed to exactly describe the electronic dynamics in this case (see Appendix~\ref{sec:Rabi}).
\footnote{In practice, a number of different algorithms have been developed to propagate these equations of motion in a more numerically stable way.\cite{Granucci2001,Barbatti2014newtonX,Mai2018SHARC}
These approaches, originally developed for the surface-hopping methods, can also be used directly in spin-LSC and MASH.}

To go beyond the classical path approximation, one would like to evolve the nuclear phase-space variables consistently with the electronic dynamics, using something akin to Newton's equation of motion. %
How to do this is the key question explored in this and many previous papers. In particular, what is the best expression to use for the force?
Formally, within the kinematic picture, the state-independent and state-dependent contributions to the nuclear force operator are given by $\bar{\mathcal{F}}_{j}(\bm{q})=-{\partial\bar{V}(\bm{q})}/{\partial q_{j}}$ and $\hat{\mathcal{F}}_{j}(\bm{q})=-{\partial\hat{V}(\bm{q})}/{\partial q_{j}}$ respectively. %
Using Eq.~(\ref{eq:V_ad}), the state-dependent nuclear force operator can be written in the adiabatic basis as
\begin{align}
\label{eq:nuclear_force}
\hat{\mathcal{F}}_{j}(\bm{q})&=-\frac{\partial V_{z}(\bm{q})}{\partial q_{j}}\hat{\sigma}_{z}(\bm{q})+2V_{z}(\bm{q})d_{j}(\bm{q})\hat{\sigma}_{x}(\bm{q}) .
\end{align}
The state-independent nuclear force is an ordinary function, but the state-dependent contribution is an electronic operator and classical-trajectory techniques differ by how they incorporate it into the theory.\cite{Stock2005nonadiabatic} Before introducing our new method in Sec.~\ref{sec:MASH}, we first consider popular pre-existing approaches.
\subsection{Fewest-Switches Surface Hopping}
In fewest-switches surface hopping (FSSH),\cite{Tully1990hopping,Subotnik2016review,Wang2016} %
the state-dependent nuclear force is given by
\begin{equation}
\label{eq:force_FSSH}
\mathcal{F}_{j}^{\text{FSSH}}(\bm{q},n_{\text{active}})=-\frac{\partial V_{z}(\bm{q})}{\partial q_{j}}\,n_{\text{active}} ,
\end{equation}
which is combined with the state-independent term, $\bar{\mathcal{F}}_{j}(\bm{q})$, to obtain the total force.
Thus, $n_{\text{active}}$ determines the active surface on which the nuclei are evolved: $n_{\text{active}}=1$ corresponds to the upper adiabatic surface and $n_{\text{active}}=-1$ corresponds to the lower one.

In order to describe nonadiabatic transitions, the active surface changes stochastically during the evolution of each trajectory with a transition probability that mimics the electronic dynamics of Eq.~(\ref{eq:mapping_eom_z}). During a time-step, $\delta t$, the probability of hopping from active surface $n_{\text{active}}=\pm1$ to $n_{\text{active}}=\mp1$ is given by
\begin{equation}
\pm\frac{2\sum_{j}d_{j}(\bm{q})p_{j}}{m}\frac{S_{x}\delta t}{1\pm S_{z}} .
\end{equation}
Negative probabilities are set to zero and the time-step must be chosen to be small enough so that the probability is never greater than one. 

To ensure that the energy function,
\begin{equation}
\label{eq:energy_FSSH}
E(\bm{p},\bm{q},n_{\text{active}})=\sum_{j=1}^{f}\frac{p_{j}^{2}}{2m}+\bar{V}(\bm{q})+V_{z}(\bm{q})\,n_{\text{active}} ,
\end{equation}
is conserved during a hop between active surfaces,
the momentum must be rescaled. It is generally agreed (although not always implemented in practice)\cite{Mai2018SHARC} that this momentum rescaling should be performed along the direction of the nonadiabatic coupling vector, $d_{j}(\bm{q})$.\cite{Tully1990hopping} As a result of this rescaling, a hop from $n_{\text{active}}=-1$ to $n_{\text{active}}=1$ can only occur if there is sufficient nuclear kinetic energy. Hops for which this is not the case must be aborted, which are commonly referred to as frustrated hops. Although there is no consensus on whether or not to alter the nuclear momentum when frustrated hops occur,\cite{Hammes-Schiffer1994,Muller1997} or whether to do so only in certain situations,\cite{Jasper2003,Jain2015hopping2} in this paper we follow the common practice of reversing the sign of the momentum along the direction of the nonadiabatic coupling vector in every case (which was Tully's original suggestion).\cite{Hammes-Schiffer1994} 

To calculate correlation functions of the form given by Eq.~(\ref{eq:corr_function}) with FSSH, the initial values of $\bm{S}$ must first be determined. For the case that the correlation function initializes the electronic subsystem in a pure electronic state, $\ket{\alpha}=\cos{(\alpha_{\theta}/2)}\eu{-i\alpha_{\phi}/2}\ket{\psi_+(\bm{q})}+\sin{(\alpha_{\theta}/2)}\eu{i\alpha_{\phi}/2}\ket{\psi_-(\bm{q})}$, it can be shown from Eq.~(\ref{eq:density_elem}) that $\bm{S}(0)=\bm{z}_{\alpha}$, where
\begin{equation}
\label{eq:FSSH_sampling}
\bm{z}_{\alpha}=\{\sin{\alpha_{\theta}}\cos{\alpha_{\phi}},\,\sin{\alpha_{\theta}}\sin{\alpha_{\phi}},\,\cos{\alpha_{\theta}}\} , 
\end{equation}
which corresponds to a single point on the Bloch sphere, with the latitude and the longitude specified by two angles $\alpha_\theta\in[0,\pi]$ and $\alpha_\phi\in[0,2\pi)$. These are the same initial conditions for the spin-coordinates used in Ehrenfest dynamics. Additionally the active surface variable, $n_{\text{active}}$, is initialized randomly so that its ensemble average over many trajectories is equal to $S_{z}(0)$. In other words, $n_{\text{active}}(0)=1$ with probability $\cos^{2}{(\alpha_{\theta}/2)}$ and $n_{\text{active}}(0)=-1$ with probability $\sin^{2}(\alpha_{\theta}/2)$. Finally, we represent the electronic contribution to the observable operator, $\hat{B}$, using the following measurement prescription\cite{Landry2013}
\begin{subequations}
\label{eq:density_FSSH}
\begin{align}
\hat{\sigma}_{x}(\bm{q})&\mapsto S_{x} , \\
\hat{\sigma}_{y}(\bm{q})&\mapsto S_{y} , \\
\hat{P}_{\pm}(\bm{q})&\mapsto \tfrac{1}{2}\left(1\pm n_{\text{active}}\right) ,
\end{align}
\end{subequations}
which is known as the density-matrix approach for calculating dynamical observables within FSSH\@. Here, we have used the notation, $\hat{P}_\pm(\bm{q})=\half[\hat{\mathcal{I}} \pm \hat{\sigma}_z(\bm{q})]$, for the population of the upper or lower adiabatic state.
\subsection{Spin Mapping}\label{sec:spin}
In contrast to FSSH, spin mapping\cite{spinmap,multispin} offers a rigorous framework for obtaining classical-trajectory approaches from approximations to the QCLE\@. For two-level subsystems, this mapping uses the fact that the electronic dynamics are analogous to a classical \mbox{spin-$\tfrac{1}{2}$} that precesses around a magnetic field. Defining the effective time-dependent magnetic field as $\bm{H}=\{0,2\sum_{j}d_{j}(\bm{q})p_{j}/m,2V_{z}(\bm{q})\}$, so that the spin system is isomorphic to the electronic subsystem considered in this paper, one finds that the resulting spin-coordinate equations of motion, $\dot{\bm{S}}=\bm{H}\times\bm{S}$, are identical to those given by Eq.~(\ref{eq:mapping_eom}). Not only does this give an additional interpretation to these equations of motion, but it also demonstrates that spin mapping can be constructed to exactly reproduce the bare electronic dynamics along a pre-specified nuclear path.

The so-called Stratonovich--Weyl transform employs the kernel $\hat{w}(\bm{S})$ to %
represent electronic operators in terms of spin-coordinate functions\cite{stratonovich1957}
\begin{subequations}
\label{eq:op_rep}
\begin{align}
B(\bm{S})&=\tr\!\left[\hat{B}\hat{w}(\bm{S})\right] , \\
\hat{w}(\bm{S})&=\tfrac{1}{2}\big[1+r_{\text{s}}\bm{S}\cdot\hat{\bm{\sigma}}(\bm{q})\big] ,
\end{align}
\end{subequations}
where $r_{\text{s}}$ is the spin-sphere radius, included explicitly here as a scaling factor in this expression so that $\bm{S}$ remains a spin-vector with unit length, in contrast to the notation of Ref.~\onlinecite{spinmap}. Note that for any value for the spin-sphere radius, the identity operator is always correctly represented by the mapping, i.e., $\mathcal{I}(\bm{S})=1$.

The dynamics of the spin-mapping trajectories are determined so as to conserve the energy function
\begin{equation}
\label{eq:energy_LSC}
E(\bm{p},\bm{q},\bm{S})=\sum_{j}\frac{p_{j}^{2}}{2m}+\bar{V}(\bm{q})+r_{\text{s}}V_{z}(\bm{q})S_{z} ,
\end{equation}
which is generated from the energy operator [Eq.~(\ref{eq:ham})] using the Stratonovich--Weyl kernel [Eq.~(\ref{eq:op_rep})]. This leads to the following expression for the state-dependent nuclear force
\begin{equation}
\label{eq:force_LSC}
\mathcal{F}_{j}^{\text{spin}}(\bm{q},\bm{S})=r_{\text{s}}\left[-\frac{\partial V_{z}(\bm{q})}{\partial q_{j}}S_{z}+2V_{z}(\bm{q})d_{j}(\bm{q})S_{x}\right] ,
\end{equation}
which could alternatively be derived using the Stratonovich--Weyl transform of Eq.~(\ref{eq:nuclear_force}). Unlike for FSSH, this is a mean-field force that evolves the nuclei on an average of the adiabatic potential-energy surfaces, weighted according to the trajectory's value of $\bm{S}$. These equations of motion are equivalent to those of MMST mapping, because the spin-variables are related to the harmonic-oscillator coordinates and momenta via a canonical transform.\cite{Wang1999mapping,spinmap} 
In previous work, we have presented the method in terms of a diabatic basis.
Here, however, we have expressed the equations of motion in the adiabatic basis in order to facilitate comparison with FSSH.
The two formalisms are exactly equivalent as, like many other mapping methods,\cite{Meyer1979nonadiabatic,Cotton2017mapping} spin mapping is independent of the electronic representation.

While the choice of the spin-sphere radius, $r_{\text{s}}$, does not effect the ability of spin mapping to reproduce the exact bare electronic dynamics associated with the classical path approximation, it does lead to different results when the electronic and nuclear subsystems are coupled, through the expression for the state-dependent nuclear force [Eq.~(\ref{eq:force_LSC})]. One popular choice for the spin-sphere radius, $r_{\text{s}}=1$, gives the so-called Ehrenfest approach, for which the spin-coordinates are initialized from a point on the Bloch sphere in the same way as for FSSH [Eq.~(\ref{eq:FSSH_sampling})]. The Ehrenfest method is, however, known to give inaccurate results for many systems. One way to improve upon Ehrenfest is to require that spin mapping exactly reproduces the QCLE up to first-order in time, the requirements of which are outlined in Appendix \ref{sec:QCLE}. It can be shown that this is only satisfied when $r_{\text{s}}=\sqrt{3}$.\cite{multispin,spinPLDM2} This makes physical sense, as only this choice of the spin-radius is able to reproduce the correct spin-magnitude of the quantum \mbox{spin-$\tfrac{1}{2}$} particle,\cite{spinmap,Cotton2013mapping} because $\hat{\sigma}_{x}(\bm{q})^{2}+\hat{\sigma}_{y}(\bm{q})^{2}+\hat{\sigma}_{z}(\bm{q})^{2}=3\hat{\mathcal{I}}$.

\begin{figure}
\includegraphics[scale=0.7]{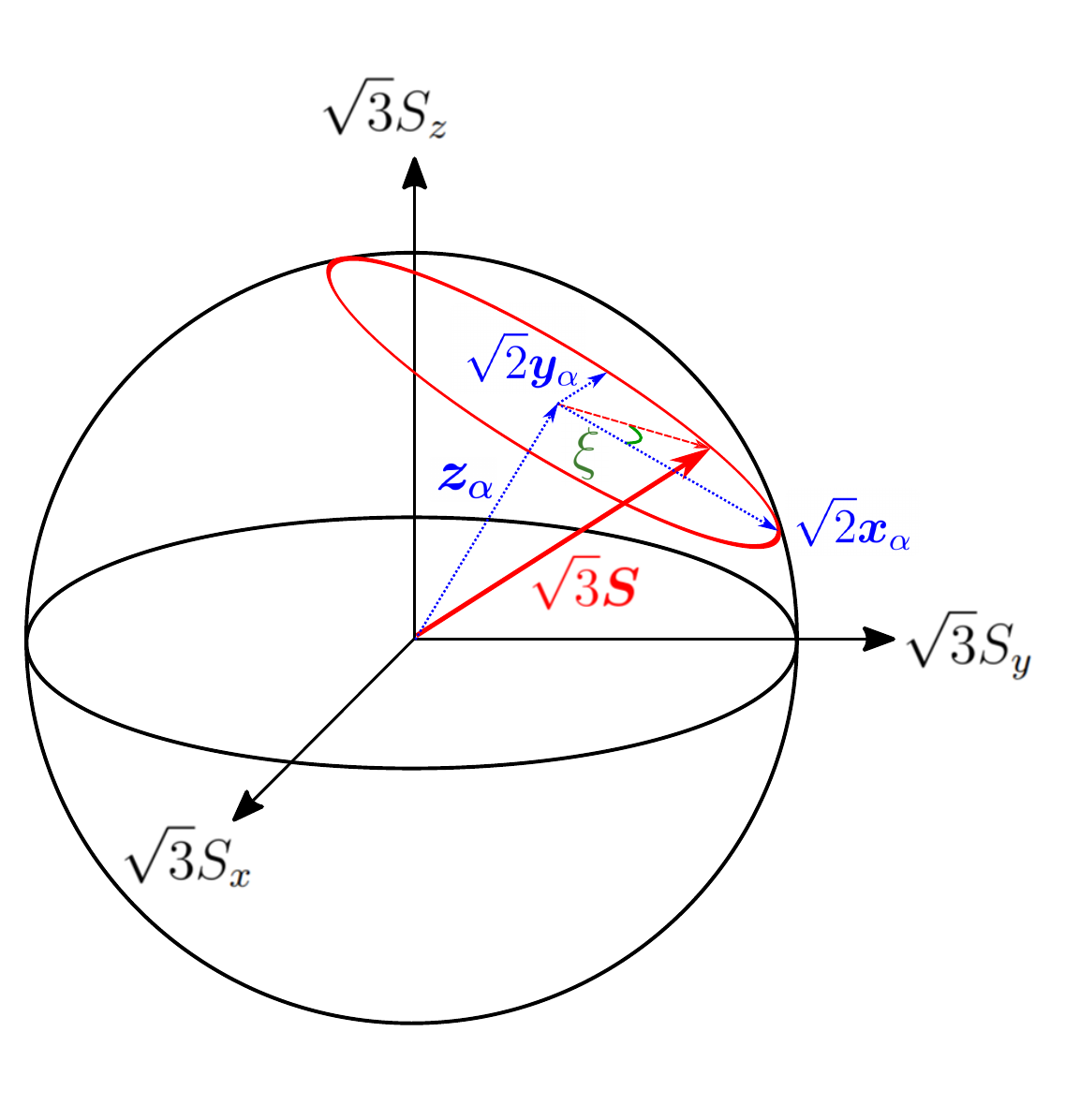}
\caption{Illustration of the spin-LSC focused initial conditions used to calculate correlation functions starting in the electronic coherent state associated with the vector $\bm{z}_{\alpha}$ on the Bloch sphere of radius one. For the spin-sphere with radius $r_{\text{s}}=\sqrt{3}$, these focused conditions correspond to sampling the spin-coordinates from the polar circle (red) centered about $\bm{z}_{\alpha}$.}\label{fig:spin_LSC}
\end{figure}
Considering again the case that the correlation function initializes the electronic subsystem in a pure electronic state, $\ket{\alpha}$,\footnote{It is also possible to initialize in a mixed state by expanding the density matrix as a sum over pure states} so-called focused initial conditions can be used, which corresponds to sampling the spin-coordinates from the region of the spin-sphere that satisfies $[\ket{\alpha}\!\bra{\alpha}](\bm{S})=1$. For $r_{\text{s}}=\sqrt{3}$, it can be shown that this region corresponds to a polar circle centered about the vector, $\bm{z}_{\alpha}$ [Eq.~(\ref{eq:FSSH_sampling})].\cite{spinmap} Such a polar circle can be defined in terms of the vector $\bm{z}_{\alpha}$ and two vectors orthogonal to it, for example $\bm{x}_{\alpha}=\{\cos{(\alpha_{\theta})}\cos{(\alpha_{\phi})},\cos{(\alpha_{\theta})}\sin{(\alpha_{\phi})},-\sin{(\alpha_{\theta})}\}$ and $\bm{y}_{\alpha}=\{-\sin{(\alpha_{\phi})},\cos{(\alpha_{\phi})},0\}$, as illustrated in Fig.~\ref{fig:spin_LSC}. The spin-vector is thus initialized as
\begin{equation}
\label{eq:samp_lsc}
\sqrt{3}\bm{S}=\bm{z}_{\alpha}+\sqrt{2}\left(\bm{x}_{\alpha}\cos\xi+\bm{y}_{\alpha}\sin\xi\right) ,
\end{equation}
where $\xi$ is an angle sampled uniformly from the range $[0,2\pi)$. Physically the sampling over the angle $\xi$ incorporates the correct quantum uncertainty in the measurement of spin operators orthogonal to the mapping representation of $\ket{\alpha}\!\bra{\alpha}$. We choose to use focused initial conditions for the spin-coordinates instead of so-called full-sphere sampling,\cite{spinmap,multispin} which corresponds to uniform sampling over the surface of the sphere, as this guarantees that the correct classical nuclear dynamics are reproduced in the Born--Oppenheimer limit when starting in a given adiabatic state. Finally the electronic contribution to the observable operator, $\hat{B}$, can also be generated from the Stratonovich--Weyl kernel [Eq.~(\ref{eq:op_rep})] to give
\begin{subequations}
\label{eq:density_LSC}
\begin{align}
\hat{\sigma}_{x}(\bm{q})&\mapsto r_{\text{s}}S_{x} , \\
\hat{\sigma}_{y}(\bm{q})&\mapsto r_{\text{s}}S_{y} , \\
\hat{P}_{\pm}(\bm{q})&\mapsto \tfrac{1}{2}\left(1\pm r_{\text{s}}S_{z}\right) .
\end{align}
\end{subequations}
As the dynamics are based on a linearized semiclassical approximation,
we will refer to the spin-mapping approach described in this section as spin-LSC.
\section{A Mapping Approach to Surface Hopping}\label{sec:MASH}
In both mean-field and surface-hopping approaches, the trajectories are constructed to conserve an energy function.
One way of distinguishing between the two types of approaches is by the form this function takes.
In mean-field approaches, like spin-LSC [Eq.~(\ref{eq:energy_LSC})], the electronic dependence enters through a continuous electronic variable,  $S_{z}$, while in surface-hopping approaches, like FSSH [Eq.~(\ref{eq:energy_FSSH})], it enters through a discrete electronic variable,  $n_{\text{active}}$. These two forms of the energy function in some sense correspond to analogue and digital representations of signals.  The predominance of the FSSH method suggests that in many cases the digital representation is superior, at least for certain types of chemical applications.

\begin{figure}
\includegraphics[scale=0.5]{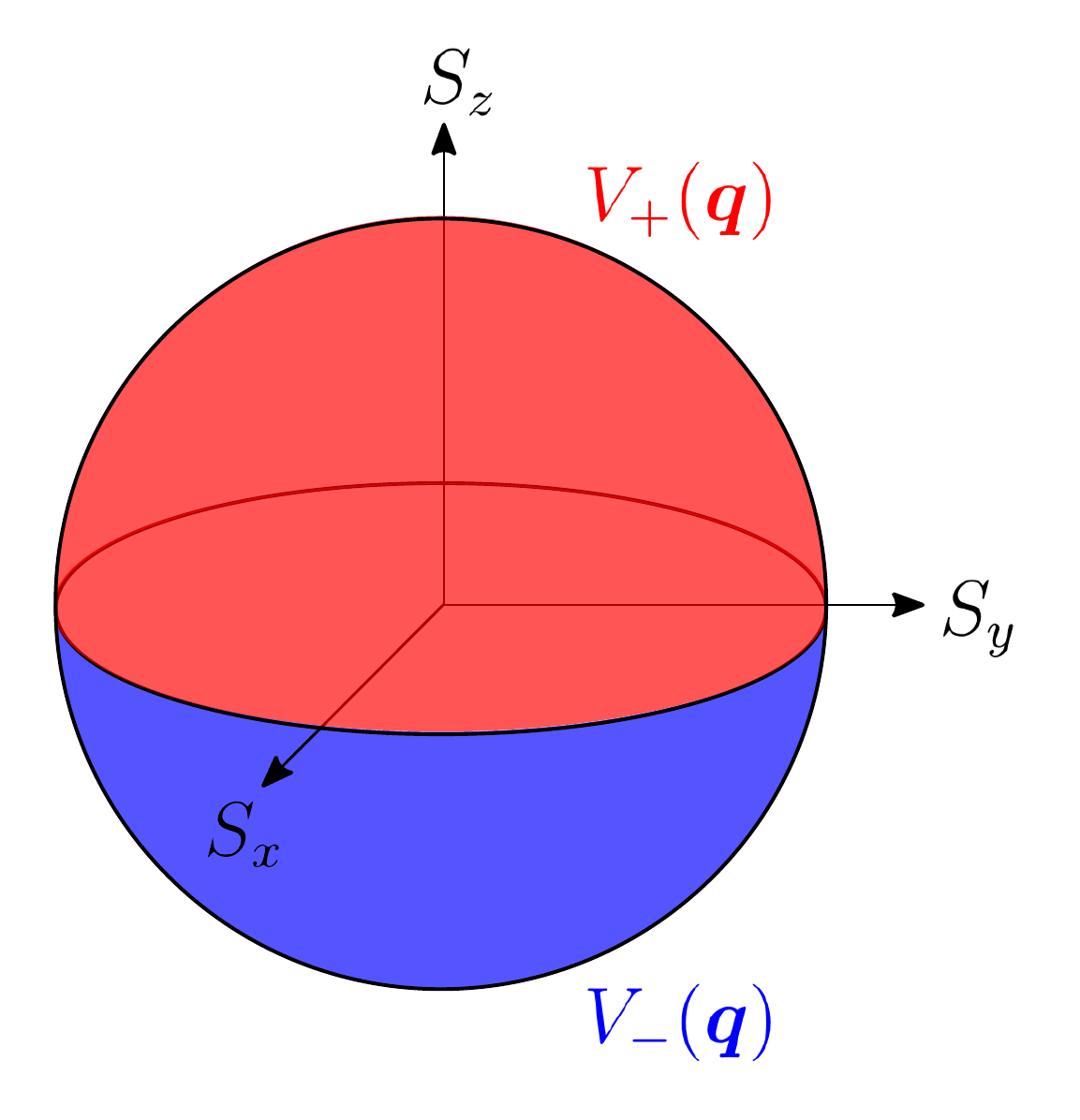}
\caption{The partitioning of the spin-sphere used within MASH to give a digital representation of the energy [Eq.~(\ref{eq:energy_MASH})]. When the spin-coordinate is in the upper hemisphere (red), the energy function returns the sum of the upper adiabatic energy, $V_{+}(\bm{q})$, and the nuclear kinetic energy and MASH propagates the nuclei on the upper adiabatic surface. In the same way, the lower hemisphere (blue) corresponds to the lower adiabatic surface, $V_{-}(\bm{q})$.}\label{fig:spin_MASH}
\end{figure}

In order to incorporate the advantages of FSSH into spin mapping, an approach for obtaining a force that propagates the nuclei on a single adiabatic surface at any time is required. Such an approach will therefore necessarily conserve an energy function whose electronic dependence enters through a discrete electronic variable; in particular the electronic contribution to this function must return a single adiabatic potential energy for each allowed value of the spin-coordinates. The most natural way of constructing this is for the energy function to return the adiabatic potential energy for which the electronic state has the highest associated probability. Thus, the function should return the upper adiabatic energy, $V_{+}(\bm{q})=\bar{V}(\bm{q})+V_{z}(\bm{q})$, when the spin-coordinate is in the northern hemisphere and the lower adiabatic energy, $V_{-}(\bm{q})=\bar{V}(\bm{q})-V_{z}(\bm{q})$, in the southern hemisphere.  This is illustrated in Fig.~\ref{fig:spin_MASH} and expressed mathematically as
\begin{equation}
\label{eq:energy_MASH}
E(\bm{p},\bm{q},\bm{S})=\sum_{j}\frac{p_{j}^{2}}{2m}+\bar{V}(\bm{q})+V_{z}(\bm{q})\,\text{sgn}\!\left(S_{z}\right) ,
\end{equation}
where $\text{sgn}(\cdot)=\pm1$ returns the sign of its argument. Because this energy function, unlike for the corresponding spin-LSC function, is invariant to the choice of the spin-sphere radius, we set $r_{\text{s}}=1$ for simplicity.

Setting the spin-coordinates to evolve under their standard equations of motion [Eq.~(\ref{eq:mapping_eom})] and the nuclear coordinates to evolve under Newton's equation of motion [%
$\dot{p}_j=\bar{\mathcal{F}}_j+\mathcal{F}^\text{MASH}_j$], an expression for the state-dependent nuclear force can be determined by requiring that the dynamics conserves this energy function (i.e., $\rd E(\bm{p},\bm{q},\bm{S})/\rd t = 0$), which gives
\begin{equation}
\label{eq:force_MASH}
\mathcal{F}^{\text{MASH}}_{j}(\bm{q},\bm{S})=-\frac{\partial V_{z}(\bm{q})}{\partial q_{j}}\text{sgn}\!\left(S_{z}\right)+4V_{z}(\bm{q})d_{j}(\bm{q})S_{x}\delta\!\left(S_{z}\right) .
\end{equation}
This force defines a new method for simulating nonadiabatic dynamics that we call a mapping approach to surface hopping (MASH). We first note that this expression for the force has a similar form to the exact state-dependent nuclear force operator [Eq.~(\ref{eq:nuclear_force})], i.e., $\hat{\sigma}_{z}(\bm{q})$ is represented by $\text{sgn}(S_{z})$ and $\hat{\sigma}_{x}(\bm{q})$ is represented by $2S_{x}\delta(S_{z})$. Indeed, we show in Appendix \ref{sec:QCLE} that with this force, the QCLE result for correlation functions can be reproduced up to first-order in time, as for spin-LSC\@. It is because of this that we are able to introduce a rigorous jump procedure in Sec.~\ref{sec:jump} to systematically recover the results of the QCLE\@. While, based solely on this criterion, Eq.~(\ref{eq:force_MASH}) is an equally justifiable expression for the force as that of spin-LSC [Eq.~(\ref{eq:force_LSC})], we expect that overall the MASH force will be advantageous as it will give rise to better underlying nuclear dynamics synonymous with surface-hopping schemes, especially in the case of wavepacket branching.
Therefore compared to spin-LSC, we expect to obtain more accurate results with MASH in these cases and to be able to systematically converge the MASH results to the QCLE using fewer jumps.

We note that the first term on the right-hand side of Eq.~(\ref{eq:force_MASH}) has the same form as the force used in FSSH [Eq.~(\ref{eq:force_FSSH})] and so MASH also propagates nuclei on a single adiabatic surface at any given time, as desired. The active surface is, however, not determined by an additional parameter within the theory, as in FSSH, but is instead determined solely from the instantaneous coordinates of the trajectory, and in particular the hemisphere in which the spin-vector resides. %
Transitions between active surfaces are now deterministic in nature (instead of stochastic as for FSSH) and occur whenever the mapping variables cross the equator. We call these `hops' due to the similarity with FSSH.

The second term in Eq.~(\ref{eq:force_MASH}) contains a Dirac delta function of the spin-mapping variable $S_{z}$ and so contributes an impulse when the mapping variables pass through the equator of the spin-sphere. %
As we will demonstrate, this term is physically responsible for rescaling the momenta after a hop in order to guarantee
energy conservation during a transition.
In cases where the transition is forbidden, it causes the mapping variables to bounce off the equator leading to a frustrated hop.
In practice, the presence of a Dirac delta function in Eq.~(\ref{eq:force_MASH}) means that this force cannot be implemented directly as written. One could, in principle, use prelimit forms of the delta function and $\text{sgn}(S_{z})$ (i.e., a Gaussian and an error function), so that the corresponding prelimit form of the energy function is conserved by the dynamics. This would require an incredibly small time-step and is thus not a practical procedure. A much more efficient way of implementing the force (and similar to what is done in FSSH) is to apply the effect of this last term as a discontinuous change in the momentum whenever the spin-vector touches the equator. %
In order to do this, we now derive the form of the momentum rescaling described by this term.
\subsection{Momentum Rescaling}\label{sec:momentum_rescaling}
Here we focus on the trajectory at the moment in which the spin-vector approaches the equator, $S_{z}(t)\approx0$, such that the last term in Eq.~(\ref{eq:force_MASH}) dominates the force and
\begin{equation}
\label{eq:impulse}
\frac{\rd p_{j}}{\rd t}\approx4V_{z}(\bm{q})d_{j}(\bm{q})S_{x}\delta\!\left(S_{z}(t)\right) .
\end{equation}
We will additionally treat %
the other phase-space variables ($\bm{q}$, $S_{x}$ and $S_{y}$) as constants for this analysis. We first note that this impulse acts in the direction of the nonadiabatic coupling vector, $d_{j}(\bm{q})$, and so only the momentum along this direction, $\tilde{p}=\sum_{j}d_{j}(\bm{q})p_{j}/\tilde{d}(\bm{q})$, will be affected, where $\tilde{d}(\bm{q})=\sqrt{\sum_{j}d_{j}(\bm{q})^{2}}$. Because the equations of motion for the mapping variables [Eq.~(\ref{eq:mapping_eom})] also only depend on the momentum along this direction, the dynamics in the vicinity of the equator can be rewritten in terms of $\tilde{p}$
\begin{subequations}
\label{eq:dynam_rescale}
\begin{align}
\label{eq:Sz_time}
&\frac{\rd S_{z}}{\rd t}=-\frac{2\tilde{p}\tilde{d}(\bm{q})S_{x}}{m} , \\
&\frac{\rd\tilde{p}}{\rd t}\approx4V_{z}(\bm{q})\tilde{d}(\bm{q})S_{x}\delta\!\left(S_{z}(t)\right) . \label{eq:impulse_tilde}
\end{align}
\end{subequations}
These two equations can be decoupled by taking the time derivative of Eq.~(\ref{eq:Sz_time}) and inserting Eq.~(\ref{eq:impulse_tilde}) to give
\begin{equation}
m\frac{\rd^{2}S_{z}}{\rd t}\approx-8V_{z}(\bm{q})\tilde{d}(\bm{q})^{2}S_{x}^{2}\delta(S_{z}(t)) ,
\end{equation}
which corresponds to Newton's equation of motion for a fictitious one-dimensional particle with position, $S_{z}$, and mass, $m$, evolving on the sign potential, $E_\mathrm{s}\,\text{sgn}(S_{z})$, where $E_\mathrm{s}=4V_{z}(\bm{q})\tilde{d}(\bm{q})^{2}S_{x}^{2}\ge0$ is a constant. %
Once we have understood the behaviour of $S_z$, we will use Eq.~(\ref{eq:Sz_time}) to determine how the nuclear momentum is altered as a result of a hop.

\begin{figure}
\includegraphics[scale=0.52]{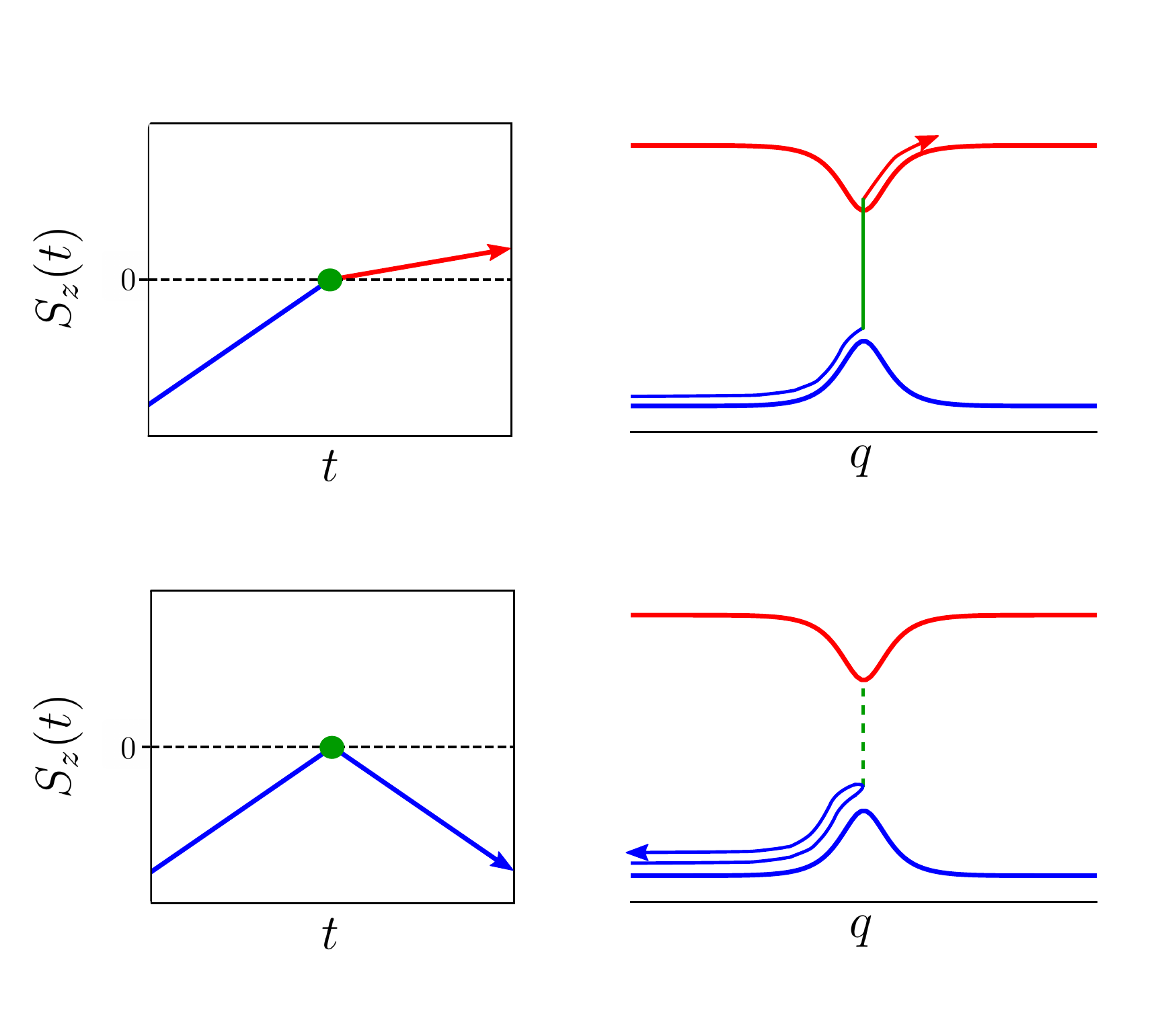}
\caption{The various possible hopping options within MASH that arise from Eqs.~(\ref{eq:dynam_rescale}) when starting in the lower adiabatic state (blue). The first column illustrates the time-dependence of $S_{z}$, while the second column shows the dynamics of the nuclear coordinate $q$. A successful hop is illustrated in the first row, where the magnitude of the momentum, $\tilde{p}$, is decreased at the hopping time in order to conserve the total energy. A successful hop from the upper to the lower adiabat would correspond to the first row in reverse. Finally the second row shows a frustrated hop that occurs for an energetically forbidden transition, for which both $S_{z}$ and $\tilde{p}$ are reflected at the hopping time.
}\label{fig:hop}
\end{figure}

There are two different situations which may arise. First, if the fictitious particle's kinetic energy, $\half m\dot{S}_{z}^{2}$, is large enough to overcome the barrier associated with the sign potential, then the spin-coordinate, $S_z$, will cross the equator (corresponding to a successful hop). This is always satisfied on going from the upper to the lower spin hemisphere (as the sign potential is barrierless in this direction), while going from the lower to the upper hemisphere can only occur if $\half m\dot{S}_{z}^{2}\geq 2E_\mathrm{s}$.
Using Eq.~(\ref{eq:Sz_time}), this condition can also be written in terms of the nuclear momentum, $\tilde{p}_{\text{init}}$, just before the hop:
$\tilde{p}_{\text{init}}^{2}/2m\geq 2V_{z}(\bm{q})$, i.e., there must be sufficient nuclear kinetic energy to compensate for the excitation energy of the electronic subsystem. Therefore, the nuclear momentum after the hop, $\tilde{p}_{\text{fin}}$, can be determined from how $\dot{S}_{z}$ changes as a result of passing through the sign potential: 
$\half m\dot{S}_{z,\text{fin}}^{2}=\half m\dot{S}_{z,\text{init}}^{2}+2E_\mathrm{s}\,\text{sgn}(\tilde{p}_{\text{init}}S_{x})$,
with $\text{sgn}(\tilde{p}_{\text{init}}S_{x})$ determining whether the electronic subsystem is excited or de-excited through Eq.~(\ref{eq:Sz_time}).
This leads to 
\begin{equation}
\label{eq:energy_conserve}
\frac{\tilde{p}_{\text{fin}}^{2}}{2m}=\frac{\tilde{p}_{\text{init}}^{2}}{2m}+2V_{z}(\bm{q})\,\text{sgn}\left(\tilde{p}_{\text{init}}S_{x}\right) ,
\end{equation}
which corresponds to a rescaling of the nuclear momentum in order to conserve the total energy of the system during the hop. The sign of the nuclear momentum is unchanged during a successful hop, because the sign of $\dot{S}_{z}$ is also unchanged.

Alternatively, if $\half m\dot{S}_{z}^{2}<2E_\mathrm{s}$, or equivalently $\tilde{p}_{\text{init}}^{2}/2m<2V_{z}(\bm{q})$, for a hop that corresponds to an excitation of the electronic subsystem, then the $S_{z}$ spin-coordinate will elastically reflect off the sign potential, causing the hop to be aborted. This therefore corresponds to a so-called frustrated hop, which occurs when there is insufficient nuclear kinetic energy to ensure energy conservation during the hop. During a frustrated hop, Eq.~(\ref{eq:Sz_time}) shows that due to the elastic reflection of $S_{z}$ (i.e., $S_{z,\text{fin}}=-S_{z,\text{init}}$), $\tilde{p}$ is also elastically reflected (i.e., $\tilde{p}_{\text{fin}}=-\tilde{p}_{\text{init}}$). %

Together these two cases constitute a unique prescription for momentum rescaling.  This is illustrated schematically in Fig.~\ref{fig:hop}. We have confirmed numerically that the dynamics indeed follows this behaviour when using the full MASH force [Eq.~(\ref{eq:force_MASH})] with consistent prelimit forms for the functions $\delta(S_{z})$ and $\text{sgn}(S_{z})$.

Performing this momentum rescaling in a discontinuous fashion at the hopping times gives rise to an efficient scheme for implementing the MASH equations of motion. Therefore in the working equations, the state-dependent nuclear force is taken to be
\begin{equation} \label{eq:short_force_MASH}
\mathcal{F}_{j}^{\text{MASH}}(\bm{q},\bm{S})=-\frac{\partial V_{z}(\bm{q})}{\partial q_{j}}\text{sgn}\!\left(S_{z}\right) ,
\end{equation}
and the effect of the last term on the right-hand side of Eq.~(\ref{eq:force_MASH}) is accounted for by performing the momentum rescaling at each hop as detailed above. %
Note that, as for FSSH, one only needs to compute the complete nonadiabatic coupling vector whenever a hop or frustrated hop is encountered. For the majority of the trajectory, only the projection of the nonadiabatic coupling vector onto the momentum is required, which can be significantly computationally cheaper if finite difference schemes are used to evaluate the derivatives within an \emph{ab initio} trajectory approach.\cite{Granucci2001}

This momentum rescaling determined from the MASH equation of motion [Eq.~(\ref{eq:impulse})] is identical to one of the standard procedures which have been proposed for FSSH\@.\cite{Hammes-Schiffer1994}
However, in FSSH, momentum rescaling is introduced in an \textit{ad hoc} way in order to ensure energy conservation.  This is not sufficient to uniquely determine the procedure and leads to various alternative suggestions as to the direction in which to rescale,\cite{Mai2018SHARC} which root of a quadratic equation to select, \cite{Muller1997}
and whether to reverse the momentum at a frustrated hop or not.\cite{Hammes-Schiffer1994,Muller1997,Jasper2003,Jain2015hopping2}
In MASH, the prescription is uniquely defined as it arises naturally from the underlying equations of motion.
Our analysis, thus, acts as a rigorous justification for why this constitutes the correct momentum rescaling, %
as the total state-dependent force given by Eq.~(\ref{eq:force_MASH}) has been generated so that its associated dynamics are guaranteed to correctly reproduce the QCLE result to at least first-order in time, as proven in Appendix \ref{sec:QCLE}.
For instance, it is necessary to employ this prescription for momentum rescaling in order that the quantum-jump procedure of Sec.~\ref{sec:jump} can be relied upon to systematically improve the results.
It is interesting to note that this finding contrasts the conclusion of Martens, who also claims that momentum rescalings in FSSH are \emph{ad hoc} and proposed that they should not be employed at all in his variant of the (stochastic) surface-hopping approach.\cite{Martens2019}
In this work, we show that one can make momentum rescalings rigorous after all, at least within the MASH formalism.
Perhaps the reason why some authors have suggested alternatives to the standard momentum-rescaling procedure can be identified with the internal inconsistency between the active state and the spin vector in FSSH.\cite{Fang1999}
Because of the internal consistency in MASH %
we expect that the problems identified with the standard momentum-rescaling procedure will be resolved.

\subsection{Correlation Functions}
Now that the equations of motion for MASH have been determined, we next consider how to evaluate correlation functions of the form given by Eq.~(\ref{eq:corr_function}) with this approach.
It is desirable to have internal consistency between how the observable operators and the nuclear forces measure the electronic subsystem. For example, when the nuclei are evolving on the upper adiabatic surface, we want the observable to show that the electronic state has a zero probability of being in the lower adiabatic state and vice versa. This is the concept behind the density-matrix approach for calculating observables in FSSH [Eq.~(\ref{eq:density_FSSH})].
In a similar way, for MASH, this requires that the observable operators are represented as
\begin{subequations}
\label{eq:density_MASH}
\begin{align}
\hat{\sigma}_{x}(\bm{q})&\mapsto S_{x} , \\
\hat{\sigma}_{y}(\bm{q})&\mapsto S_{y} , \\
\hat{P}_{\pm}(\bm{q})&\mapsto h(\pm S_{z}) , 
\end{align}
\end{subequations}
where $h(x)$ is the Heaviside step function defined so that the function is in keeping with its prelimit form (i.e., $h(x)=1$ for $x>0$, $h(x)=\tfrac{1}{2}$ for $x=0$ and $h(x)=0$ for $x<0$); $h(\pm S_{z})$ thus reports on whether the spin-vector is in the upper or lower hemisphere. %
A measurement of $\hat{P}_\pm(\bm{q})$ will therefore only select trajectories in the corresponding hemisphere.
Because of this, trajectories will be initialized with the appropriate active surface when starting in an adiabatic population, and MASH, like FSSH, is guaranteed to reduce to classical nuclear dynamics on a single surface in the Born--Oppenheimer limit ($d_j(\bm{q})\rightarrow0$). 

Using these operator representations, correlation functions can be calculated with MASH using the following expression
\begin{equation}
\label{eq:corr_MASH}
C_{AB}(t)\approx\Big\langle A(\bm{q},\bm{p},\bm{S})\,\mathcal{W}_{AB}(\bm{S})\,B(\bm{q}(t),\bm{p}(t),\bm{S}(t))\Big\rangle ,
\end{equation}
where $\langle\cdots\rangle$ is defined by Eq.~(\ref{eq:av_def}). Rather than sampling the spin-coordinates using focused initial conditions, as is done for FSSH and Ehrenfest, MASH instead uses full-sphere sampling, where the spin-coordinates are uniformly sampled from the surface of the spin-sphere. This is necessary for a number of reasons, one of which is that in order to enable a transition from one spin-hemisphere to the other in a system with weak nonadiabatic coupling, some trajectories must be initialized close to the equator.
The same reasoning is behind why it is necessary to use touching windows within SQC.\cite{Cotton2016SQC} For full-sphere sampling, the integral over the spin-coordinates is defined by
\begin{equation}
\label{eq:spin_int}
\int\rd\bm{S}\cdots=\frac{1}{2\pi}\int_{0}^{\pi}\rd\theta\sin\theta\int_{0}^{2\pi}\rd\phi\cdots ,
\end{equation}
where $S_{x}=\sin{\theta}\cos{\phi}$, $S_{y}=\sin{\theta}\sin{\phi}$ and $S_{z}=\cos{\theta}$. This integral is typically evaluated by Monte Carlo (along with the integral over the nuclear variables) by sampling $\cos\theta$ and $\phi$ uniformly from the ranges $[-1,1]$ and $[0,2\pi)$ respectively.

The MASH correlation function expression differs subtly from the usual phase-space expression [Eq.~(\ref{eq:corr_phase})] because Eq.~(\ref{eq:corr_MASH}) contains a positive-definite weighting factor, given by
\begin{equation}
\label{eq:MASH_weight}
\mathcal{W}_{AB}(\bm{S})=\begin{cases}
	3 \quad&\text{if}\,\, AB=\text{CC}, \\
	2 \quad&\text{if}\,\, AB=\text{CP}\,\, \text{or}\,\, \text{PC},\\
	2|S_{z}| \quad&\text{if}\,\, AB=\text{PP},
	\end{cases}
\end{equation}
where the weight depends on whether the operators are coherences, $\mathrm{C}\in\{\hat{\sigma}_{x}(\bm{q}),\hat{\sigma}_{y}(\bm{q})\}$, or populations, $\mathrm{P}\in\{\hat{P}_{+}(\bm{q}),\hat{P}_{-}(\bm{q})\}$. These weighting factors have been constructed in order to guarantee that MASH not only exactly
describes the bare electronic dynamics along a pre-specified nuclear path (as shown in Appendix \ref{sec:Rabi}) but also exactly reproduces the QCLE dynamics up to at least first-order in time (as shown in Appendix \ref{sec:QCLE}).

Equations \eqref{eq:short_force_MASH}--\eqref{eq:MASH_weight} are the main result of this work and define the standard MASH method.
We next turn to error estimates and approaches for correcting any problems identified.

\subsection{Microscopic Reversibility Error}

While the weighting factors for the CC, CP or PC correlation functions [Eq.~\ref{eq:MASH_weight}] are simply constants,
for PP correlation functions, the associated weighting factor $\mathcal{W}_{\text{PP}}(\bm{S})$ is in general time-dependent. This means that, as for many other nonadiabatic trajectory methods\cite{ellipsoid} including FSSH,\cite{Jain2015hopping1} microscopic reversibility is not rigorously obeyed by MASH (as discussed further in Appendix~\ref{sec:detailed_balance}).
Because of this, we had a choice as to whether the weighting-factor should be applied at time $t=0$ or at time $t$. Both choices give the same results in all the cases where MASH is exact and therefore both choices are by this metric equally justified. Throughout this work, we will choose to apply the weighting at time $t=0$, as is consistent with Eq.~\eqref{eq:corr_MASH}, so that the sum of the adiabatic populations is guaranteed to be equal to one. 

However, the fact that there is a non-unique choice in the weighting leads to the following definition of a microscopic reversibility error (MRE) 
\begin{equation}
\label{eq:error_function}
\Big\langle A(\bm{q},\bm{p},\bm{S})\,\left[\mathcal{W}_{AB}(\bm{S}(t))-\mathcal{W}_{AB}(\bm{S})\right]\,B(\bm{q}(t),\bm{p}(t),\bm{S}(t))\Big\rangle ,
\end{equation}
which is zero in all the cases where MASH exactly reproduces the QCLE (see Appendix \ref{sec:Rabi}).
It can be very useful to have a simple measure of the error of a method.
However, note that it only reports on the breakdown of microscopic reversibility
and not on other potential deficiencies of the dynamics,
such as zero-point energy leakage or the incorrect description of other quantum nuclear effects. 

A significant deviation of this quantity from zero signifies that MASH is no longer a good approximation to the QCLE\@. %
To solve this problem, we will now explain how the accuracy of the MASH results can be systematically improved by resampling the spin-coordinates at intermediate times within a quantum-jump procedure, similar to those which have been successfully implemented with other mapping approaches.\cite{Hsieh2013FBTS,Huo2012PLDM,spinPLDM2} %
We then show how this can be used as a rigorous starting point for deriving an approximate decoherence correction.
\subsection{Quantum-Jump Procedure}\label{sec:jump}
So far, we have used trajectory approaches to evolve $\hat{B}$ forwards in time and measure its correlation with $\hat{A}$ at time zero (equivalent to the Heisenberg picture).
There is of course an equivalent framework (the Schr\"odinger picture) in which one propagates the initial density matrix $\hat{\rho}(\bm{q},\bm{p},t=0)=\hat{A}(\bm{q},\bm{p})$ forwards in time to give $\hat{\rho}(\bm{q}_{t},\bm{p}_{t},t)$, where $\{\bm{q}_t,\bm{p}_t\}$ are dummy phase-space variables. %
Within MASH, this time-evolved density matrix is given by
\begin{subequations}
\label{eq:MASH_densitymatrix}
\begin{align}
\begin{split}
&\hat{\rho}(\bm{q}_{t},\bm{p}_{t},t)\approx \\
&\int\rd\bm{q}\rd\bm{p}\int\rd\bm{S}\, \delta(\bm{q}_{t}-\bm{q}(t))\delta(\bm{p}_{t}-\bm{p}(t))A(\bm{q},\bm{p},\bm{S})\hat{\mathcal{W}}(\bm{S},t)
\end{split} \\
&\hat{\mathcal{W}}(\bm{S},t)=\begin{pmatrix}
\mathcal{W}_{A\text{P}}(\bm{S})\,h(S_{z}(t)) & \mathcal{W}_{A\text{C}}(\bm{S})\frac{S_{x}(t)-\iu S_{y}(t)}{2} \\
\mathcal{W}_{A\text{C}}(\bm{S})\frac{S_{x}(t)+\iu S_{y}(t)}{2} & \mathcal{W}_{A\text{P}}(\bm{S})\,h(-S_{z}(t)) 
\end{pmatrix} , \label{eq:MASH_densitymatrix2}
\end{align}
\end{subequations}
from which the MASH correlation functions [Eqs.~(\ref{eq:density_MASH}) and (\ref{eq:corr_MASH})] can be generated using $C_{\text{AB}}(t)\approx\int\rd\bm{q}_{t}\rd\bm{p}_{t}\,\tr[\hat{\rho}(\bm{q}_{t},\bm{p}_{t},t)\hat{B}(\bm{q}_{t},\bm{p}_{t})]/(2\pi)^{f}$. 

We know that MASH is accurate for short times, but that the error may increase at longer times.
In order to avoid this error, one can
first propagate the density matrix [Eq.~(\ref{eq:MASH_densitymatrix})] for a shorter time-interval, $t_{0}$, after initially sampling a set of phase-space variables, $\{\bm{q},\bm{p},\bm{S}^{(0)}\}$. At the end of this time-interval, the density matrix associated with each trajectory is reinitialized by setting the new initial observable operator to $A(\bm{q},\bm{p},\bm{S}^{(0)})\hat{\mathcal{W}}(\bm{S}^{(0)},t_{0})$ and evolving the density matrix for the next time-interval, $t_{1}$, using Eq.~(\ref{eq:MASH_densitymatrix}) with a resampled set of spin-mapping variables, $\bm{S}^{(1)}$. This is often referred to as a jump,\cite{Hsieh2013FBTS,Huo2012PLDM,spinPLDM2}
and it can be carried out $m$ times by
selecting the time-intervals such that they add up to give the total time, $t=t_{0}+\cdots+t_{m}$.
Note that the nuclear variables are not resampled, because we want to treat the nuclear dynamics entirely classically in accordance with the QCLE\@.

\begin{figure*}
\includegraphics[scale=0.32]{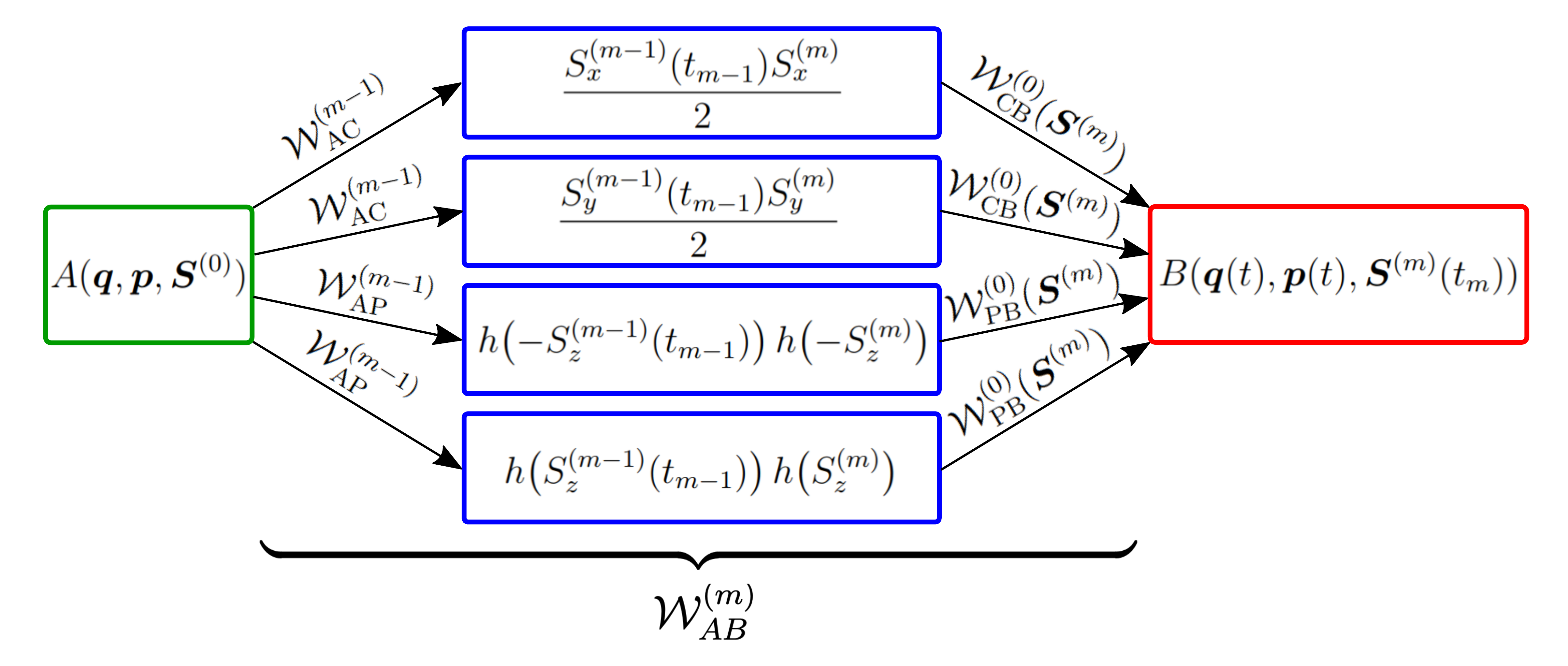}
\caption{A schematic illustrating the MASH $C_{AB}(t)$ correlation function with $m$ jumps, which %
defines a recursive relation for the weighting factors used in the quantum-jump procedure and the approximate decoherence correction. In this diagram, the first $m-1$ jumps are included implicitly through the weighting factors $\mathcal{W}^{(m-1)}_{A\text{C}}$ and $\mathcal{W}^{(m-1)}_{A\text{P}}$, while the final jump is explicitly described by considering all possible intermediate electronic density-matrix elements that the subsystem can reach by the end of the penultimate time-interval $t_{m-1}$ (i.e., $\hat{\sigma}_{x}(\bm{q})$, $\hat{\sigma}_{y}(\bm{q})$, $\hat{P}_{-}(\bm{q})$ and $\hat{P}_{+}(\bm{q})$, each of which corresponds to a blue box in the diagram). The overall diagram therefore corresponds to $\langle A(\bm{q},\bm{p},\bm{S}^{(0)})\,\mathcal{W}^{(m)}_{AB}\,B(\bm{q}(t),\bm{p}(t),\bm{S}^{(m)}(t_{m}))\rangle$. The weighting factor for $m$ jumps, $\mathcal{W}^{(m)}_{AB}$, can therefore be obtained by summing over the contributions from the four pathways. The analogous weighting factor for the simpler decoherence correction corresponds to summing over just the two lower pathways. 
}\label{fig:jump}
\end{figure*}
The definition of the MASH correlation function after $m$ jumps is obtained with a slight modification of the expression given in Eq.~(\ref{eq:corr_MASH}).
The operator functions become $A(\bm{q},\bm{p},\bm{S}^{(0)})$ and $B(\bm{q}(t),\bm{p}(t),\bm{S}^{(m)}(t_{m}))$, where $\bm{S}^{(0)}$ are the mapping variables at the start of the first time interval and $\bm{S}^{(m)}(t_m)$ are the mapping variables at the end of the final time-interval. %
The weighting factor $\mathcal{W}_{AB}^{(m)}$ will depend on all of the resampled mapping variables. The expression for $\mathcal{W}_{AB}^{(m)}$ can be constructed in a recursive fashion using the schematic diagram in Fig.~\ref{fig:jump}.
The weighting factor for the final jump is given as a sum over four terms, which depend on the weighting factors obtained from the previous $m-1$ jumps.
The four terms correspond to a decomposition of the time-evolved density matrix [Eq.~(\ref{eq:MASH_densitymatrix})]
into the components  $\{\hat{\sigma}_{x}(\bm{q}),\hat{\sigma}_{y}(\bm{q}),\hat{P}_{-}(\bm{q}),\hat{P}_{+}(\bm{q})\}$
and depend on 
the value of the mapping variables just before the final jump, $\bm{S}^{(m-1)}(t_{m-1})$. %
These terms are then multiplied by the MASH functions [Eq.~(\ref{eq:density_MASH})] containing the resampled mapping variable, $\bm{S}^{(m)}$.
These resampled variables are associated with the standard (no-jump) weighting factors, $\mathcal{W}^{(0)}_{AB}(\bm{S}^{(m)})$, given by Eq.~(\ref{eq:MASH_weight}).
Finally, they are propagated over the final time-interval to give $\bm{S}^{(m)}(t_{m})$ and are used to measure $B$.
This recursion relation can be formalised in mathematical notation as:
\begin{subequations}
\label{eq:jump}
\begin{align}
\label{eq:jump_weight1}
\begin{split}
\mathcal{W}^{(m)}_{A\mathrm{C}}=&\frac{3}{2}\left[S^{(m-1)}_{x}(t_{m-1})S^{(m)}_{x}+S^{(m-1)}_{y}(t_{m-1})S^{(m)}_{y}\right]\mathcal{W}^{(m-1)}_{A\mathrm{C}}\\
&+\left[1+\text{sgn}\!\left(S^{(m-1)}_{z}(t_{m-1})S^{(m)}_{z}\right)\right]\mathcal{W}^{(m-1)}_{A\mathrm{P}} ,
\end{split} \\ 
\label{eq:jump_weight2}
\begin{split}
\mathcal{W}^{(m)}_{A\mathrm{P}}=&\left[S^{(m-1)}_{x}(t_{m-1})S^{(m)}_{x}+S^{(m-1)}_{y}(t_{m-1})S^{(m)}_{y}\right]\mathcal{W}^{(m-1)}_{A\mathrm{C}} \\
&+|S^{(m)}_{z}|\left[1+\text{sgn}\!\left(S^{(m-1)}_{z}(t_{m-1})S^{(m)}_{z}\right)\right]\mathcal{W}^{(m-1)}_{A\mathrm{P}} ,
\end{split}
\end{align}
\end{subequations}
where we have also used $h(-x)h(-y)+h(x)h(y)=\half[{1+\text{sgn}(xy)}]$, i.e., this factor is only non-zero when $x$ and $y$ have the same sign. %
Notice that in general these weighting factors are non-zero for all values of the spin-coordinates and so every jump resamples the mapping variables from all regions of the spin-sphere surface.

In order to calculate the MRE through Eq.~(\ref{eq:error_function}) when applying jumps, an associated expression for $\mathcal{W}^{(m)}_{AB}(t)$ is needed. This can be obtained through a modified recursive formula, where at each level of the hierarchy,
the weight at the start of each interval is replaced by the weight at the end, i.e., $|S^{(m)}_{z}|$ is replaced with $|S_{z}^{(m)}(t_{m})|$ in Eq.~(\ref{eq:jump_weight2}).

Because Eq.~(\ref{eq:MASH_densitymatrix}) reproduces the QCLE result up to at least first-order in time (Appendix~\ref{sec:QCLE}), MASH is guaranteed to reproduce the QCLE if jumps are applied often enough. However, the statistical error %
significantly increases with the number of jumps, and so only a few jumps can ever be performed for a given simulation without dramatically increasing the computational expense. Nonetheless, it has often been observed in practice that only a few jumps are sufficient to essentially converge the accuracy of results to that of the QCLE\@.\cite{spinPLDM2,Kelly2013GQME}
An important point is that the jump procedure cannot destroy the quality of the results, unlike the FSSH decoherence schemes, because the resampling is carried out in a way that means the electronic coherences are also resampled at the jump and are not artificially set to zero. This scheme is also formally advantageous compared to commonly used FSSH decoherence schemes, as it can in principle introduce effects beyond purely decoherence phenomena, such as electronic quantum interference between recombining wavepackets (as we will demonstrate in Sec.~\ref{sec:tully2}).
\subsection{Decoherence Correction}
The quantum-jump scheme given by Eq.~(\ref{eq:jump}) can also act as a rigorous starting point from which approximate decoherence schemes can be derived within MASH\@. Considering a time $t$ that the system has correctly decohered, the ensemble average of the coherence variables should satisfy $\braket{S_{x}(t)}=\braket{S_{y}(t)}=0$, for every possible value of the nuclear phase-space variables. In this situation, the first term on the right-hand sides of Eqs.~(\ref{eq:jump_weight1}) and (\ref{eq:jump_weight2}) can be set to zero, resulting in the following approximate decoherence correction%
\footnote{In an analogous fashion to the case of the full quantum-jump procedure, the weighting factors $\mathcal{W}^{m}_{AB}(t)$ required for the MRE can be obtained from Eq.~(\ref{eq:decoherence2}) by replacing $|S_{z}^{(m)}|$ with $|S_{z}^{(m)}(t_{m})|$.}
\begin{subequations}
\label{eq:decoherence}
\begin{align}
\mathcal{W}^{(m)}_{A\mathrm{C}}&\approx\left[1+\text{sgn}\!\left(S^{(m-1)}_{z}(t_{m-1})S^{(m)}_{z}\right)\right]\mathcal{W}^{(m-1)}_{A\mathrm{P}} , \\ 
\mathcal{W}^{(m)}_{A\mathrm{P}}&\approx|S^{(m)}_{z}|\left[1+\text{sgn}\!\left(S^{(m-1)}_{z}(t_{m-1})S^{(m)}_{z}\right)\right] \mathcal{W}^{(m-1)}_{A\mathrm{P}}.
\label{eq:decoherence2}
\end{align}
\end{subequations}
The weighting factor is only non-zero when the new mapping variables are sampled from the hemisphere associated with the current active surface.
This is analogous to the decoherence corrections used within FSSH (which reset the spin vector to the pole corresponding to the active surface).\cite{Hammes-Schiffer1994} One important difference is that the MASH decoherence correction is not \emph{ad hoc} in nature but has the more rigorous justification presented above. Of course, this formal derivation relies on the assumption that it is only applied at points where $\braket{S_x(t)}=\braket{S_y(t)}=0$ and thus %
it would lead to severe errors in the dynamics if it is applied at the wrong time.
It cannot therefore be relied upon to systematically converge to the QCLE in general.
However, these decoherence corrections can be far more efficient that the quantum jumps (i.e., require a smaller value of $m$ and thus fewer trajectories in order to fully correct the results) in cases where the error in the MASH dynamics purely arises due to an incorrect description of decoherence.

\section{Results}\label{sec:results}
In this section we apply MASH to a variety of different model systems, in order to ascertain its accuracy in comparison to other commonly used classical-trajectory techniques. For many two-level models, the state-dependent potential is most easily defined in the diabatic basis as follows
\begin{equation}
\label{eq:ham_diab}
\hat{V}(\bm{q})=\kappa(\bm{q})\hat{\sigma}^{\text{diab}}_{z}+\Delta(\bm{q})\hat{\sigma}^{\text{diab}}_{x} .
\end{equation}
Expressions for the adiabatic  potential-energy surfaces, $V_{\pm}(\bm{q})=\bar{V}(\bm{q})\pm V_{z}(\bm{q})$, and the nonadiabatic coupling vectors, $d_{j}(\bm{q})$, can be obtained analytically in terms of $\Delta(\bm{q})$ and $\kappa(\bm{q})$, as shown in Appendix~\ref{sec:transf}. These expressions can then be used to compute the nuclear forces for MASH and FSSH\@. In future work, the MASH method will be applied with \emph{ab initio} electronic structure, in which case the adiabatic states are obtained directly and do not require a transform. 

For each of the following models, the system is initialized in a product state
\begin{equation}
\hat{A}=\rho_{0}(\bm{q},\bm{p}) \ket{\alpha}\!\bra{\alpha} ,
\end{equation}
where $\rho_{0}(\bm{q},\bm{p})$ is the initial nuclear density function and $\ket{\alpha}$ is an electronic state. We start by considering the three scattering models introduced by Tully before tackling spin--boson models and the internal conversion of pyrazine.
\subsection{The Tully models}
The Tully models\cite{Tully1990hopping} are well known benchmark scattering problems that offer an extensive test of the ability of an approach to simultaneously describe both the electronic and nuclear dynamics.  In particular, this includes decoherence and recoherence phenomena associated with the splitting and recombination of nuclear wavepackets. Historically, they have been standard tests for FSSH and their variants.  Each of these models contains a single nuclear degree of freedom with mass, $m=2000$, given in atomic units.

The electronic subsystem is initialized in the lower adiabatic state. For FSSH and Ehrenfest, this corresponds to initializing the spin-coordinates on the south pole [Eq.~(\ref{eq:FSSH_sampling}) with $\bm{z}_{\alpha}=\{0,0,-1\}$], while for spin-LSC, these coordinates are initialized from the polar circle centered at this pole [Eq.~(\ref{eq:samp_lsc})]. The electronic observables are then obtained for FSSH using Eq.~(\ref{eq:density_FSSH}) and for Ehrenfest and spin-LSC using Eq.~(\ref{eq:density_LSC}), with $r_{\text{s}}=1$ and $\sqrt{3}$ respectively. In contrast, MASH samples the initial spin-coordinates from the whole lower hemisphere, according to the definition of the correlation function given by Eqs.~(\ref{eq:density_MASH}) and (\ref{eq:corr_MASH}).

In some cases, the problem is defined in terms of an initial wavepacket,
centred at $\bar{q}$ with average momentum $\bar{p}$ and width $\gamma$,
for which numerically exact results are computed using a split-operator approach.\cite{Tannor}
The classical-trajectory methods are initialized using the equivalent
Wigner distribution:
\begin{equation}
\label{eq:wavepacket}
\rho_{\text{0}}(q,p)=2\exp\!\left[-\frac{1}{\gamma}(p-\bar{p})^{2}-\gamma(q-\bar{q})^{2}\right] .
\end{equation}
Other cases are defined in terms of scattering probabilities,
for which numerically exact results are computed using the log-derivative approach.\cite{Mano1986logderivative,Alexander1989logderivative}
Here, the classical-trajectory approaches are initialized with a single position and momentum, $\rho_{\text{0}}(q,p)=2\pi\delta{(q-q_{\text{init}})}\,\delta{(p-p_{\text{init}})}$.
In all cases, we use $\bar{q}=q_{\text{init}}=-15$.

One of the most important features of the Tully models is that they induce branching of the nuclear wavepackets on passing through a nonadiabatic crossing. This is particularly difficult for mean-field approaches to capture, as they evolve the nuclei on an average potential-energy surface and so typically give rise to a single peak in their associated nuclear phase-space distributions.\cite{Miller2009mapping} This branching has however been successfully incorporated into mean-field mapping approaches through the description of the electronic subsystem using a `spin polymer',\cite{SPMD} the semiclassical initial-value representation (SCIVR)\cite{Sun1997mapping,Ananth2007SCIVR,Miller2009mapping,Miller2012coherence} and methods based on coupled trajectories.\cite{Min2015nonadiabatic,Agostini2016,Agostini2018,Ibele2020} As we will see, MASH provides a completely different solution to this problem, which builds on the success of surface-hopping approaches for these models,\cite{Tully1990hopping,Subotnik2011AFSSH} but within a rigorous framework.
\subsubsection{Tully's model I}\label{sec:tully1}
The simplest of Tully's models is the single avoided crossing (model I), which tests whether an approach can correctly describe the branching of a nuclear wavepacket on passing through a region of nonadiabatic coupling. We choose to use a slightly modified version of this model proposed by Miller and coworkers\cite{Ananth2007SCIVR}
\begin{subequations}
\begin{align}
\kappa(q)&=A\tanh{\left(Bq\right)} , \\
\Delta(q)&=C\,\eu{-Dq^{2}} , 
\end{align}
\end{subequations}
where $A=0.01$, $B=1.6$, $C=0.005$ and $D=1$.

\begin{figure}
\includegraphics[scale=0.7]{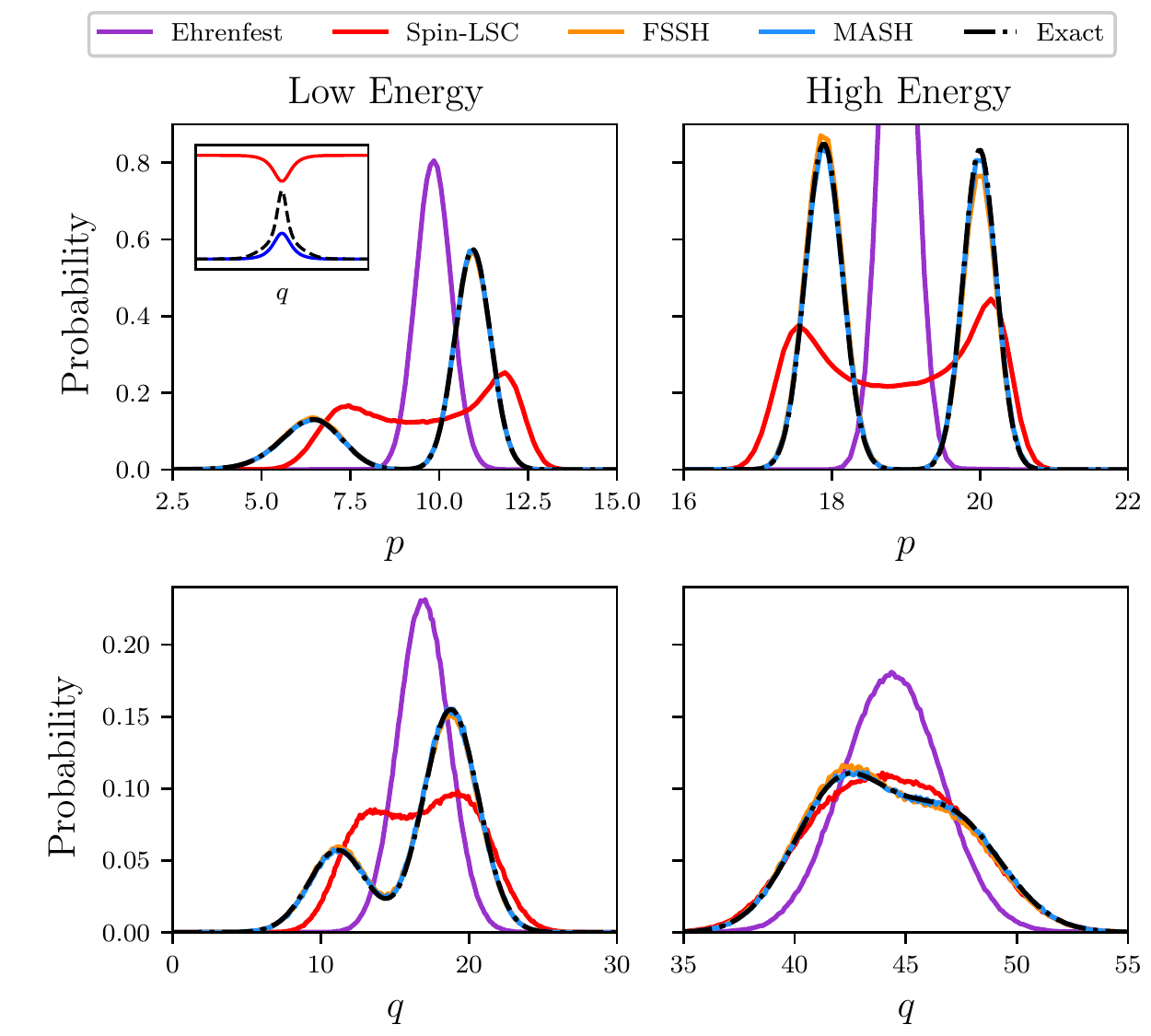}
\caption{Probability distributions for the nuclear momentum (first row) and position (second row) calculated at $t=150$\,fs for a modified version of Tully's model I\@.\cite{Ananth2007SCIVR}
Note that our spin-LSC results differ from Ref.~\onlinecite{SPMD} because here we use focused initial conditions to sample the spin-coordinates. Additionally the inset in the top left figure plots the upper and lower adiabatic surfaces (red and blue lines respectively) and the nonadiabatic coupling vector (dashed line, divided by 120).}\label{fig:tully1}
\end{figure}

Both the adiabatic potential energies and the nonadiabatic coupling vector for this model are shown in the inset of Fig.~\ref{fig:tully1}. The wavepacket is initialized according to Eq.~(\ref{eq:wavepacket}) on the lower adiabatic surface to the left of the crossing region with a positive average initial momentum. When the dynamics pass through the crossing region, there is some nonadiabatic population transfer to the upper surface. Due to energy conservation, the resulting wavepacket on the upper surface moves more slowly than the one on the lower surface, eventually resulting in the wavepacket splitting and giving rise to two peaks in the nuclear position and momentum distributions. 

These probability distributions at $t=150$ fs are presented in Fig.~\ref{fig:tully1}, for both a low- and high-energy initial wavepacket (where $\bar{p}^{2}/2m=0.03$, $\gamma=0.5$ and $\bar{p}^{2}/2m=0.1$, $\gamma=0.1$ respectively, in accordance with Ref.~\onlinecite{SPMD}). For classical-trajectory approaches, the probability distributions are computed by histogramming the associated time-evolved nuclear phase-space variables. Ehrenfest dynamics incorrectly produces a single peak in both the position and momentum distributions, which arises because the nuclear degree of freedom is evolved on a weighted average of the adiabatic surfaces and so the splitting of the nuclear wavepacket cannot be described at all. While spin-LSC does remarkably well for a mean-field method by producing two peaks in the distributions, they are, however, not located in the correct positions and are unphysically broadened. In contrast, both FSSH and MASH correctly describe the splitting of the nuclear wavepackets, due to the fact that both methods evolve the nuclear degree of freedom on a single adiabatic surface at all times. %
Tully's model I therefore illustrates that for the standard case where FSSH works well, MASH also inherits the correct behaviour.
\subsubsection{Tully's model II}\label{sec:tully2}
Tully's dual avoided crossing (model II) tests how well a method can describe electronic quantum interference effects and is significantly more difficult than model I\@. The Hamiltonian is defined by
\begin{subequations}
\begin{align}
\kappa(q)&=-\bar{V}(q)=\tfrac{1}{2}\left(A\eu{-Bq^{2}}-E_{0}\right) , \\
\Delta(q)&=C\,\eu{-Dq^{2}} , 
\end{align}
\end{subequations}
where $A=0.1$, $B=0.28$, $C=0.015$, $D=0.06$ and $E_{0}=0.05$.

\begin{figure}
\includegraphics[scale=0.9]{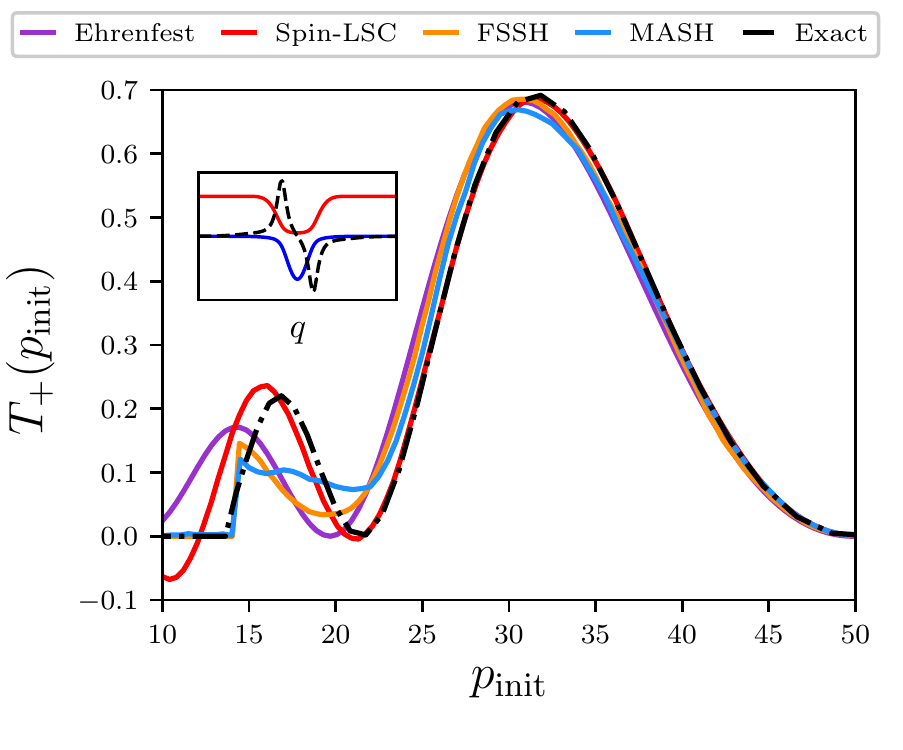}
\caption{The scattering probability associated with transmission on the upper adiabatic state as a function of the initial momentum, $p_{\text{init}}$, for Tully's model II\@. Our Ehrenfest %
and FSSH %
results are in agreement with those already published,\cite{Subotnik2011AFSSH,SPMD} whereas our spin-LSC results differ slightly from Ref.~\onlinecite{SPMD} because we use focused initial conditions to sample the spin-coordinates. Additionally the inset plots the upper and lower adiabatic surfaces (red and blue lines respectively) and the nonadiabatic coupling vector (dashed line, divided by 12).}\label{fig:tully2}
\end{figure}

The adiabatic potential energies and the nonadiabatic coupling vector are shown in the inset of Fig.~\ref{fig:tully2}. There are two avoided crossings. The system is initialized to the left of the first crossing, on the lower surface (blue) and with a single positive momentum, $p_{\text{init}}$. Transmission on the upper surface can occur via two different routes, for which either the upper or lower adiabatic state is predominantly occupied in the region between the two crossings. Each of these routes has associated with it a different momentum-dependent phase, giving rise to constructive or destructive interference when these routes recombine at the second avoided crossing. So-called St{\"u}ckelberg oscillations are observed in the scattering probability associated with transmission on the upper adiabatic state as a result of this momentum-dependent electronic interference. %

As shown in Fig.~\ref{fig:tully2}, for large initial momenta (${p_{\text{init}}\gtrsim20}$), spin-LSC is the best approach for reproducing the correct transmission probability, for which it is essentially numerically exact. This is because the mean-field force is well suited for accurately describing the highly coherent dynamics that occurs between the two crossings and the positively and negatively weighted regions of the spin-LSC sphere allow the method to better describe quantum interference effects in the electronic degrees of freedom compared to the Ehrenfest sphere. %
Although not perfect, other mean-field approaches like Ehrenfest and SQC\cite{Cotton2013mapping,Miller2016Faraday} (not shown) still perform quite well here. In particular, they correctly predict zero transmission on the upper surface for $p_{\text{init}}\approx 20$. In contrast, for low initial momenta ($p_{\text{init}}\lesssim15$), only FSSH and MASH correctly predict a zero transmission probability on the upper surface. This is because their trajectories conserve energy and evolve the nuclei on a single adiabatic surface at any given time, meaning that the upper surface is never populated when there is insufficient energy to do so. In the intermediate momentum range ($15\lesssim p_{\text{init}}\lesssim20$), all methods perform poorly.

\begin{figure}
\includegraphics[scale=0.7]{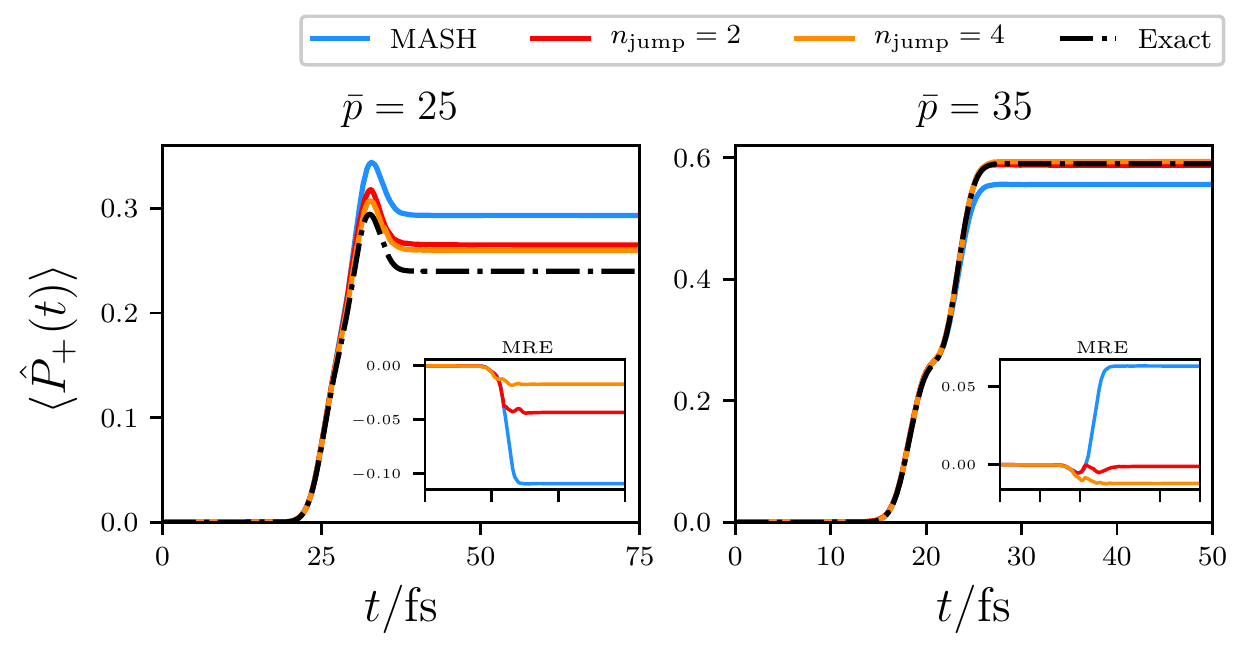}
\caption{The time-dependent adiabatic populations associated with the wavepacket dynamics of Tully's model II, with different average initial momenta, $\bar{p}$. The nuclear distribution function for the initial wavepacket is given by Eq.~(\ref{eq:wavepacket}), with $\gamma=0.5$. The insets give the MRE for these populations, defined by Eq.~(\ref{eq:error_function}).
}\label{fig:tully2_packet}
\end{figure}
To improve the accuracy of the obtained transmission probabilities with MASH, a way of incorporating both the correct nuclear motion associated with coherent electronic dynamics and negatively weighted trajectories is needed for describing electronic interference effects. The inability of surface-hopping approaches to describe decoherence is not a problem in this case, as confirmed by the fact that decoherence corrections to FSSH do not noticeably improve the accuracy of the results.\cite{Subotnik2011AFSSH} Because the jump scheme within MASH allows the active state to change when the spin-coordinates are resampled and gives rise to new weighting factors that can be both positive and negative, this scheme should, in principle, be able to include these missing effects when applied frequently enough in the region between the two crossings.

We return to a time-dependent wavepacket setup and focus only on a couple of initial average momenta in order to clearly represent the results.
Figure~\ref{fig:tully2_packet} presents the results of the MASH jump procedure, for which 2 or 4 jumps were applied during the time the trajectories are found between the two crossings. For $\bar{p}=25$, the first jump is applied at $t_0=19.5\,\text{fs}$ and the remaining jumps are equally spaced by $T/n_\text{jump}$, where $T=20.5\,\text{fs}$ and $n_\text{jump}$ is the total number of jumps. %
Analogously, for $\bar{p}=35$, we choose $t_0=13.5\,\text{fs}$ and $T=15.5\,\text{fs}$. %
This leads to a significant improvement in the accuracy of the predicted adiabatic populations, especially in the case of $\bar{p}=35$. This improvement is also qualitatively described by the associated MREs. (Note that fortuitous error cancellation makes the $n_\text{jump}=2$ result for $\bar{p}=35$ appear better than $n_\text{jump}=4$). Although it is expected that the results for $\bar{p}=25$ will continue to improve as more jumps are added, the computational expense increases and the calculations become unfeasible before full convergence is reached.  It is also expected to be difficult to converge the jump procedure at even lower values of $\bar{p}$.
Therefore, it will be necessary to develop a more efficient but approximate procedure in future work.
A phenomenological correction has already been developed for FSSH to improve the results obtained for Tully's model II,\cite{Shenvi2011phase} and it will be interesting to see if corrections like this can be rigorously derived within the MASH scheme. %
\subsubsection{Tully's model III}\label{sec:tully3}
The extended coupling model (model III) tests how well a method can describe the decoherence that arises from nuclear wavepacket splitting after leaving a region of nonadiabatic coupling. The Hamiltonian is defined by
\begin{subequations}
\begin{align}
\kappa(q)&=-A , \\
\Delta(q)&=B\left(1+\text{sgn}(q)\left(1-\eu{-C|q|}\right)\right) , 
\end{align}
\end{subequations}
where $A=6\times10^{-4}$, $B=0.1$ and $C=0.9$.

\begin{figure}
\includegraphics[scale=0.7]{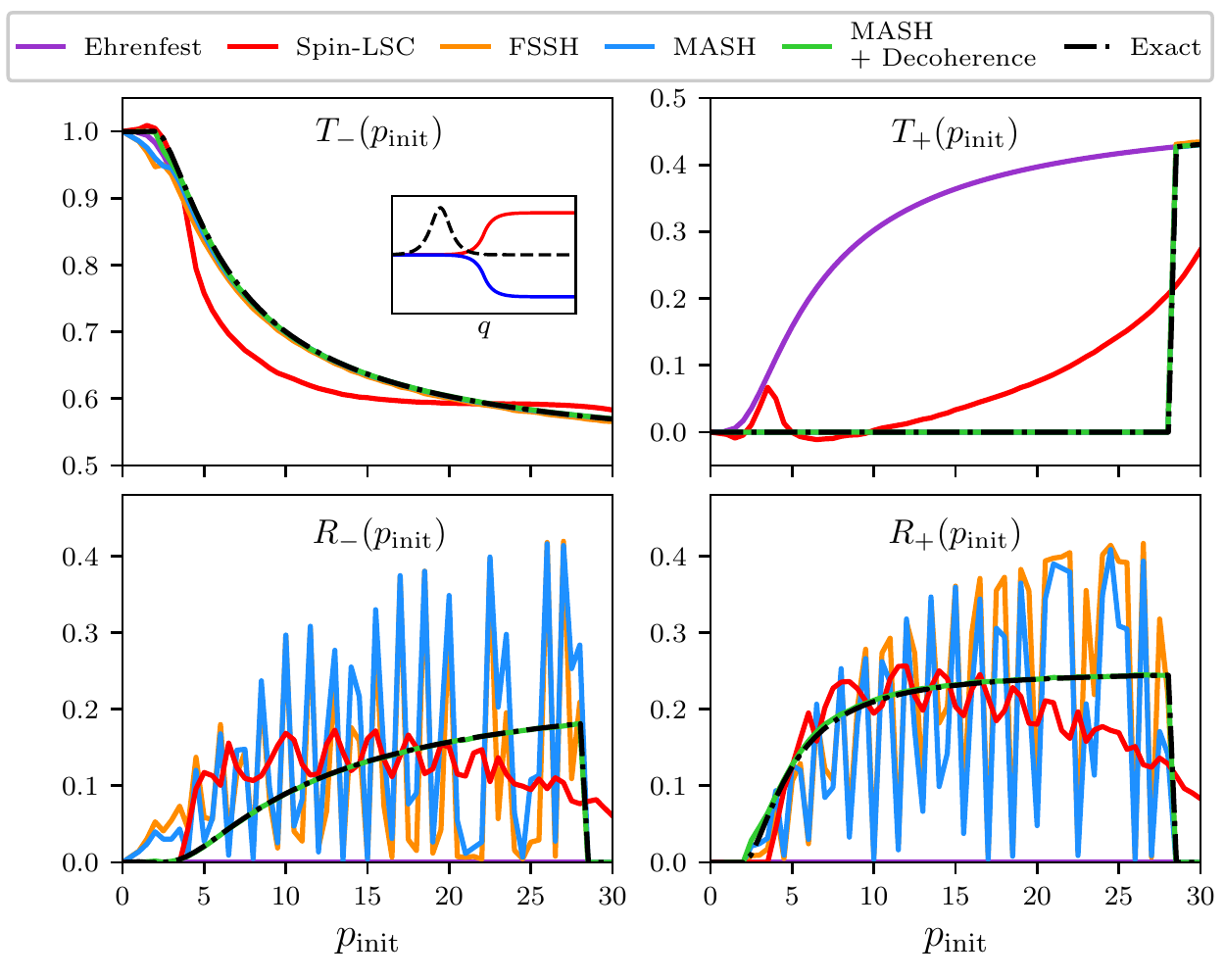}
\caption{The scattering probabilities associated with reflection ($R$) and transmission ($T$) on each adiabatic state as a function of the initial momentum, $p_{\text{init}}$, for Tully's model III\@. Our %
FSSH %
results are in agreement with those already published,\cite{Subotnik2011AFSSH} %
whereas our spin-LSC results differ from Ref.~\onlinecite{SPMD} because here we use focused initial conditions to sample the spin-coordinates. Finally the inset in the top left panel plots the adiabatic surfaces (solid lines) and the nonadiabatic coupling vector (dashed line).}\label{fig:tully3_scat}
\end{figure}
\begin{figure}
\includegraphics[scale=0.7]{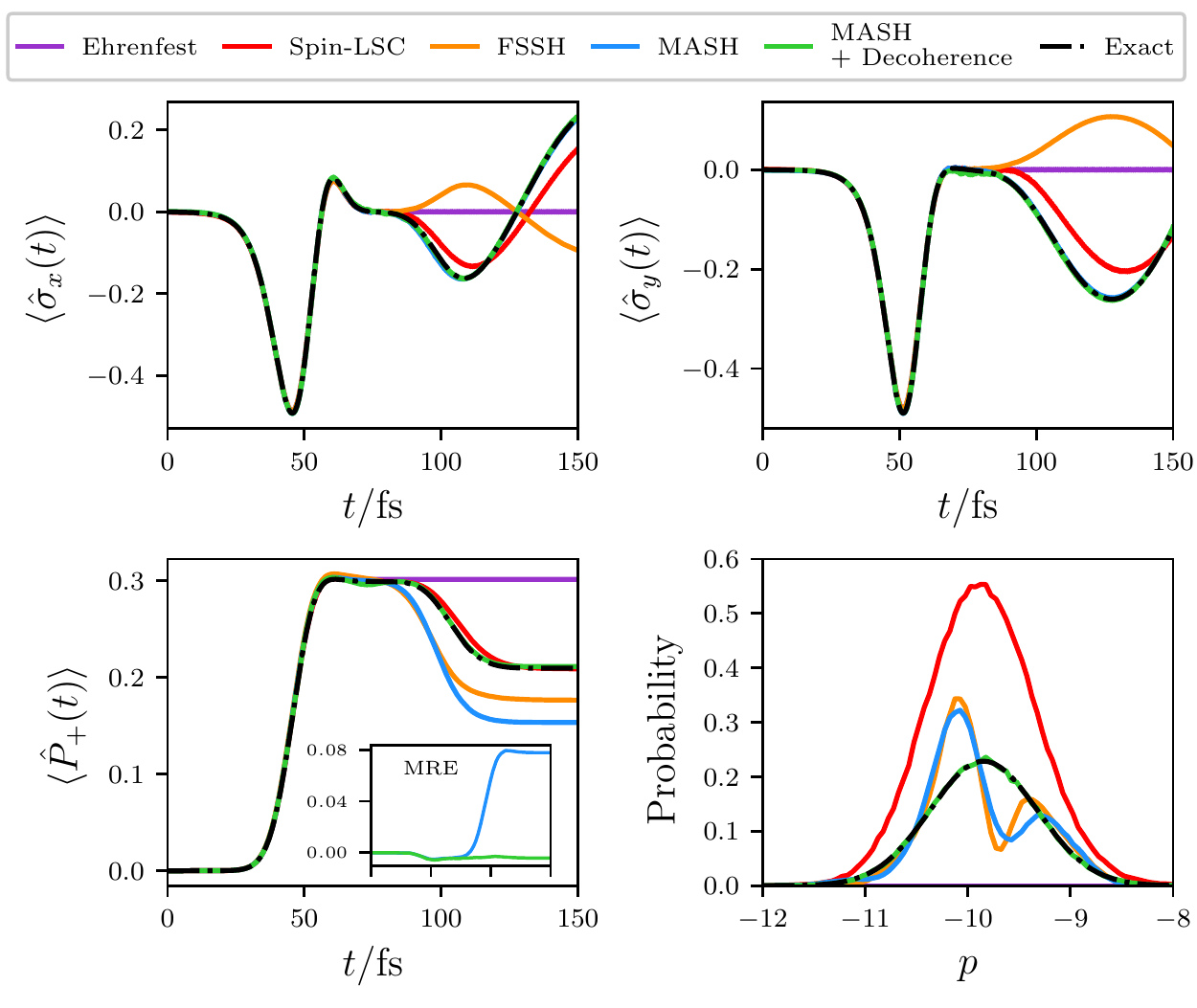}
\caption{Electronic and nuclear observables associated with the wavepacket dynamics for Tully's model III\@. The nuclear distribution function for the initial wavepacket is given by Eq.~(\ref{eq:wavepacket}), with $\bar{q}=-15$, $\bar{p}=10$ and $\gamma=0.5$, as used in Ref.~\onlinecite{SPMD}. The nuclear distribution function given by the bottom right panel shows the reflected part of the wavepacket calculated at $t=150$ fs. Finally the inset in the bottom left panel shows the MRE for the population--population correlation defined by Eq.~(\ref{eq:error_function}).
}\label{fig:tully3}
\end{figure}

The inset in the top left panel of Fig.~\ref{fig:tully3} displays the adiabatic surfaces and nonadiabatic coupling vector, which give rise to similar dynamics as Tully's model I for the first crossing of the coupling region. After this first crossing, everything on the lower surface transmits, while for $p_{\text{init}}\lesssim28$, all trajectories on the upper surface reflect and recross the coupling region. For this recrossing to be correctly described, the reflected and transmitted trajectories must have decohered from each other before the coupling region is reached again.

Figure~\ref{fig:tully3_scat} presents the momentum-dependent scattering probabilities for all possible transmission and reflection channels. For all mean-field methods when $p_{\text{init}}\lesssim28$, transmission on the energetically inaccessible upper surface, $T_{+}(p_{\text{init}})$, is incorrectly predicted to be non-zero, whereas approaches based on surface hopping can describe this by construction. Although SQC windows the observables, the forces are still mean-field, and thus it behaves similarly to spin-LSC in this case.\cite{Cotton2013mapping}
Additionally the transmission on the lower surface is also very well described by these surface-hopping approaches. This is because for single-crossing problems (like for Tully's model I) the associated nuclear and electronic dynamics are essentially numerically exact. However, at low initial momenta, $p_{\text{init}}\lesssim4$, the FSSH and MASH transmission probabilities are slightly lower than the exact result. This error occurs because both surface-hopping approaches still try and hop to the upper surface, even though the dynamics is essentially adiabatic in this case. This gives rise to frustrated hops and momentum reversal that leads to a non-zero reflection probability on the lower surface.
It may be possible to reduce this error by making \emph{ad hoc} changes to the momentum-reversal scheme.  However, this would break the connection to the QCLE and make the use of the quantum-jump procedure or decoherence corrections unjustifiable.

The main error of the surface-hopping approaches is in the reflection probabilities for larger initial momenta, which have unphysical oscillations as a result of the trajectories not properly decohering before they recross the coupling region. Although not as severe, spin-LSC also shows similar oscillations.\footnote{These oscillations are absent in the spin-LSC results presented in Ref.~\onlinecite{SPMD}, because initial conditions were used to fix the energy (according to the spin-mapping Hamiltonian) rather than the momentum.} %
Thus none of the classical-trajectory approaches can correctly describe decoherence in this case. %

These errors observed in the scattering probabilities computed using surface-hopping approaches can, however, be corrected using decoherence corrections. For MASH, we use the decoherence scheme described by Eq.~(\ref{eq:decoherence}), which is applied at most once for each trajectory. In order to correct the error at low momenta observed for transmission on the lower adiabat, the correction is applied just before the first frustrated hop occurs for each trajectory. Finally to correct the error in the reflection probabilities, the correction is applied at the exact time each trajectory reflects, i.e., when $p(t)=0$. If a trajectory has not undergone frustrated hops or was not reflected, then no decoherence correction is applied. Applying the decoherence correction in this way essentially reproduces the exact results for this problem. Similar improvements in accuracy have also been observed when decoherence corrections were applied to FSSH for this problem.\cite{Subotnik2011AFSSH} In principle, the quantum-jump scheme [Eq.~(\ref{eq:jump})] could have been used to systematically improve the MASH results, but multiple jumps would have been needed to fully converge the simulation. The decoherence correction is therefore a much more efficient way of correcting the error when only decoherence needs to be added.  

It is also of interest to study the same model initialized by a 
wavepacket [Eq.~(\ref{eq:wavepacket})], with $\bar{p}=10$ and $\gamma=0.5$.
The bottom left panel of Fig.~\ref{fig:tully3} shows the dynamical population of the upper adiabat, for which all methods are numerically exact up until the recrossing of the coupling region. The fact that both FSSH and MASH do not correctly capture the population transfer at this second transition is because averaging the reflection probabilities (Fig.~\ref{fig:tully3_scat}) over the initial momentum distribution is not guaranteed to reproduce the correct associated probabilities for the wavepacket, a point that has already been previously noted in Ref.~\onlinecite{Subotnik2011decoherence} for FSSH\@. Additionally, it is entirely coincidental that spin-LSC appears to correctly predict the population transfer at this second transition, as the overall shape of the spin-LSC reflection probability given in Fig.~\ref{fig:tully3_scat} is also not close to the exact result. The decoherence error can also be observed in the momentum distribution of the reflected wavepacket (bottom right panel), where both FSSH and MASH incorrectly predict two peaks, which arise because the reflection probabilities oscillate as a function of the momentum (Fig~\ref{fig:tully3_scat}).
It is useful to note that in MASH, the MRE (shown in the inset) clearly indicates that there is a problem at about 75\,fs.  Even if we did not know the exact result, we could use this measure to identify the problem and suggest a cure.
In particular, as for the scattering probabilities, these errors can all be corrected for by applying a decoherence correction at the point when each trajectory reflects, giving corrected MASH results which match the numerically exact ones. This improvement in the accuracy of the results is also captured by the associated MRE.

The panels in the top row of Fig.~\ref{fig:tully3} show the time-dependence of the coherences. Similar decoherence indicators were studied in Refs.~\onlinecite{SPMD} and \onlinecite{Min2015nonadiabatic}. After the second transition, FSSH predicts the wrong sign for both coherences. This arises because the reflected trajectories have an active state corresponding to the upper adiabat, but still have an electronic wavefunction which is predominately on the lower state. This inconsistency leads to an %
error in the coherences which is not observed in MASH, because the electronic and active state always remain consistent throughout the dynamics.
Therefore, this error in FSSH has a different origin from the usual ``overcoherence problem'', which leads to the errors observed in the populations.
FSSH can use decoherence corrections to fix the errors in both the populations and the coherences,
whereas MASH only requires them to fix the errors in the populations.
As there is only one type of error that has to be fixed by decoherence corrections in MASH, this may reduce the number that have to be applied and help design better and simpler algorithms for determining the optimal times to apply them.
This is the first indication that the more rigorous derivation of MASH will lead to improved results over FSSH and we will explore further cases in the following. 
\subsection{Multidimensional Models}
Tully's models are classic test cases for surface-hopping methods, whereas mapping approaches have been more commonly benchmarked on system--bath or linear vibronic-coupling models.
These multidimensional models are expected to pose a greater challenge for surface-hopping methods, as they typically give rise to many hops during the dynamics.
In contrast, mapping approaches typically behave well, especially in cases where the potentials are harmonic.
We now wish to ascertain how well MASH compares to other mapping approaches like spin-LSC when simulating nonadiabatic dynamics in these multidimensional models. 

Due to the large number of nuclear degrees of freedom, specialized numerically exact approaches for these systems can typically only compute electronic observables in the diabatic basis. Hence we will be interested in computing observables associated with the diabatic population $\hat{P}_{1}=\half[\hat{\mathcal{I}}+\hat{\sigma}^{\text{diab}}_{z}]$. For FSSH and Ehrenfest, this corresponds to initializing the spin-coordinates from a point on the Bloch-sphere [Eq.~(\ref{eq:FSSH_sampling})] given by Eq.~(\ref{eq:z_diab}), while for spin-LSC, these coordinates are initialized from the polar circle centered about this point [Eq.~(\ref{eq:samp_lsc})]. 

Expressing this diabatic population in the adiabatic basis, using Eq.~(\ref{eq:transform_z}), leads to three terms consisting of the operators $\hat{\mathcal{I}}$, $\hat{\sigma}_{x}(\bm{q})$ and $\hat{\sigma}_{z}(\bm{q})$. Using this resulting expression for the diabatic population, the corresponding observable at time $t$ can be obtained for FSSH using Eq.~(\ref{eq:density_FSSH}) and for Ehrenfest and spin-LSC using Eq.~(\ref{eq:density_LSC}), with $r_{\text{s}}=1$ and $\sqrt{3}$ respectively. For MASH, the associated observables are computed through the definition of the correlation function given in Eq.~(\ref{eq:corr_MASH}). For a diabatic population--population correlation, this gives rise to nine separate terms when expressed in the adiabatic basis, each of which must be weighted according to Eq.~(\ref{eq:MASH_weight}). These nine terms are trivial to compute all together from a single simulation, but are messy to write out explicitly. From this expression for the diabatic population--population correlation, we find that the initial spin-coordinates in MASH must therefore be sampled from the entire surface of the spin-sphere.

For the other diabatic population, $\hat{P}_{2}=\half[\hat{\mathcal{I}}-\hat{\sigma}^{\text{diab}}_{z}]$, the approach for calculating observables is analogous for each method, except now $-\bm{z}_{\alpha}$ must be used and the different signs for the terms in the expression for $\hat{P}_{2}$ in the adiabatic basis must be taken into account.
\subsubsection{Spin--Boson Models}\label{sec:spin-boson}
The prototypical model used for describing nonadiabatic dynamics in condensed-phase systems is the spin--boson model\cite{Garg1985spinboson}
\begin{subequations}
\begin{align}
\bar{V}(\bm{q})&=\tfrac{1}{2}\sum_{j=1}^{f}\omega_{j}^{2}q_{j}^{2} , \\
\kappa(\bm{q})&=\epsilon+\sum_{j=1}^{f}c_{j}q_{j} , \\
\Delta(\bm{q})&=\Delta ,
\end{align}
\end{subequations}
which consists of a two-level system linearly coupled to a harmonic bath of $f$ modes with frequencies, $\omega_{j}$. For this Hamiltonian, $\epsilon$ is the energy bias, $\Delta$ the coupling between diabatic states and $c_{j}$ the coupling strength of mode $j$ to the electronic subsystem. The nuclear frequencies, $\omega_{j}$, and the electronic--nuclear couplings, $c_{j}$, are obtained from the Debye spectral density,
\begin{equation}
\label{eq:Debye_spectral}
J(\omega)=\frac{\Lambda}{2}\frac{\omega\omega_{\text{c}}}{\omega^{2}+\omega^{2}_{\text{c}}} ,
\end{equation}
where $\Lambda$ is the Marcus reorganization energy and $\omega_{\text{c}}$ is the characteristic frequency of the bath. For our classical-trajectory simulations, we discretize this spectral density into $f=100$ modes using the scheme described in Ref.~\onlinecite{Craig2007condensed}. Additionally, the system is initialized in the first diabatic state, $\ket{1}$, where the nuclear phase-space variables are sampled from a Wigner distribution corresponding to the thermal density of the uncoupled bath,
\begin{equation}
\label{eq:bath_Wigner}
\rho_{\text{0}}(\bm{q},\bm{p})=\prod_{j=1}^{f}2\zeta_{j}\exp\!\left[-\frac{2\zeta_{j}}{\omega_{j}}\left(\frac{p_{j}^{2}}{2}+\frac{1}{2}\omega_{j}^{2}q_{j}^{2}\right)\right],
\end{equation}
where $\zeta_{j}=\tanh{(\tfrac{1}{2}\beta\omega_{j})}$ and $\beta$ is the reciprocal temperature of the bath. Numerically exact results for the associated dynamics can be computed using the hierarchical equations of motion (HEOM) approach, which we calculated using the open-source \texttt{pyrho} package.\cite{pyrho}

\begin{figure*}
\includegraphics[scale=1]{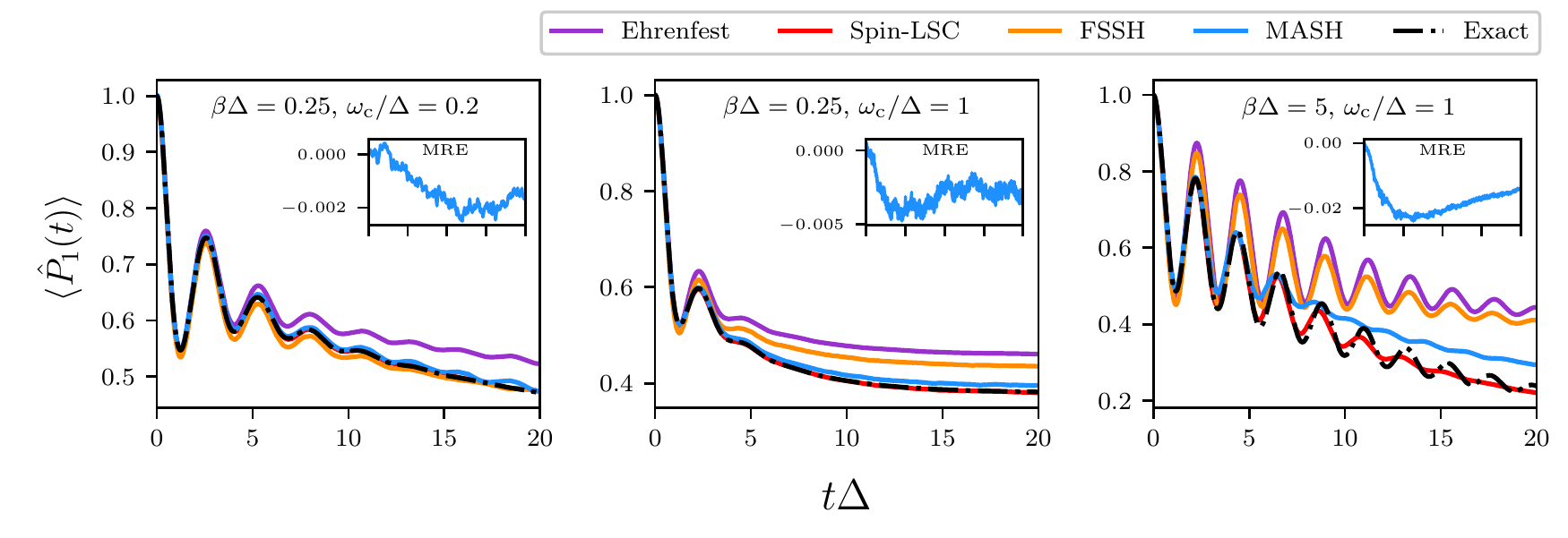}
\caption{The time-dependent diabatic populations for three spin--boson models with the Debye spectral density. The electronic subsystem was initialized in the first diabatic state, $\ket{1}$, and the nuclei from a Wigner distribution corresponding to the thermal density of the uncoupled bath [Eq.~(\ref{eq:bath_Wigner})]. All of the spin--boson models here have energy bias, $\epsilon/\Delta=1$ and Marcus reorganization energy, $\Lambda/\Delta=0.5$, while the remaining parameters are given in the respective figures. Additionally the inserts give the microscopic reversibility error (MRE) for the diabatic population--population correlation, defined according to Eq.~(\ref{eq:error_function}). The noise in the MRE is purely statistical and could in principle be removed by averaging over more trajectories.}\label{fig:spin-low}
\end{figure*}
We focus on parameter regimes that are known to be particularly challenging for FSSH\@. Figure~\ref{fig:spin-low} presents the diabatic populations for a series of spin--boson models approaching the low-temperature limit. The far right panel corresponds to the lowest temperature spin--boson model considered in this paper, for which quantum nuclear effects are significant as highlighted by the fact that initializing the nuclei with a classical distribution (instead of a Wigner distribution) drastically changes the results (not shown). Both Ehrenfest and FSSH deviate from the exact result at relatively short times and also produce a large error in the long-time limit. This is expected to be due to zero-point energy (ZPE) leakage from the nuclear degrees of freedom, which leads to a higher effective temperature of the steady-state populations. Although not perfect, MASH is able to reproduce the exact initial dynamics for longer times and also has a smaller error in the final populations than for FSSH\@. Because the QCLE is exact for spin--boson models, the quantum-jump scheme (Sec.~\ref{sec:jump}) could of course be used to further improve the accuracy of the dynamics and in principle will systematically converge to the exact result. Figure~\ref{fig:spin-low}, however, shows that spin-LSC is the method of choice for treating low-temperature spin--boson models, although its ability to correctly describe quantum nuclear effects is likely to be limited to harmonic system--bath models. As we will demonstrate in Sec.~\ref{sec:pyrazine24d}, spin-LSC also fails due to additional problems for more realistic models.

Raising the temperature of the bath (middle panel) leads to a regime where MASH can very accurately reproduce the population dynamics, while FSSH and Ehrenfest still give significant errors. The small discrepancy in the MASH result compared to the exact one can be removed by initializing the trajectories from a classical distribution in this case (not shown), which demonstrates that the error associated with ZPE leakage from a Wigner distribution is often greater for surface-hopping approaches than the error associated with completely neglecting nuclear ZPE by using a classical distribution. Finally reducing the characteristic frequency of the bath (left panel) leads to a spin--boson model in the classical-nuclear limit, for which both FSSH and MASH give a very good description of the dynamics. Note, however, that Ehrenfest can still give significant errors even in the classical-nuclear limit, as has been found in previous work.\cite{spinmap,identity,spinPLDM1}

The inset of each figure shows the MRE associated with the diabatic population--population correlation. Comparing the MRE with the difference between the MASH and exact results demonstrates that the MRE clearly does not give a quantitative description of the total error in MASH\@. This is to be expected, as the MRE does not report on the main source of error in these models, which comes from the incorrect description of quantum nuclear effects. The MRE is, however, still observed to qualitatively describe the correct trend in the actual error on changing the temperature of the bath. We therefore expect that the MRE will still provide a useful indicator of the reliability of MASH results, even when the main source of error does not arise from the violation of microscopic reversibility.

\begin{figure}
\includegraphics[scale=0.7]{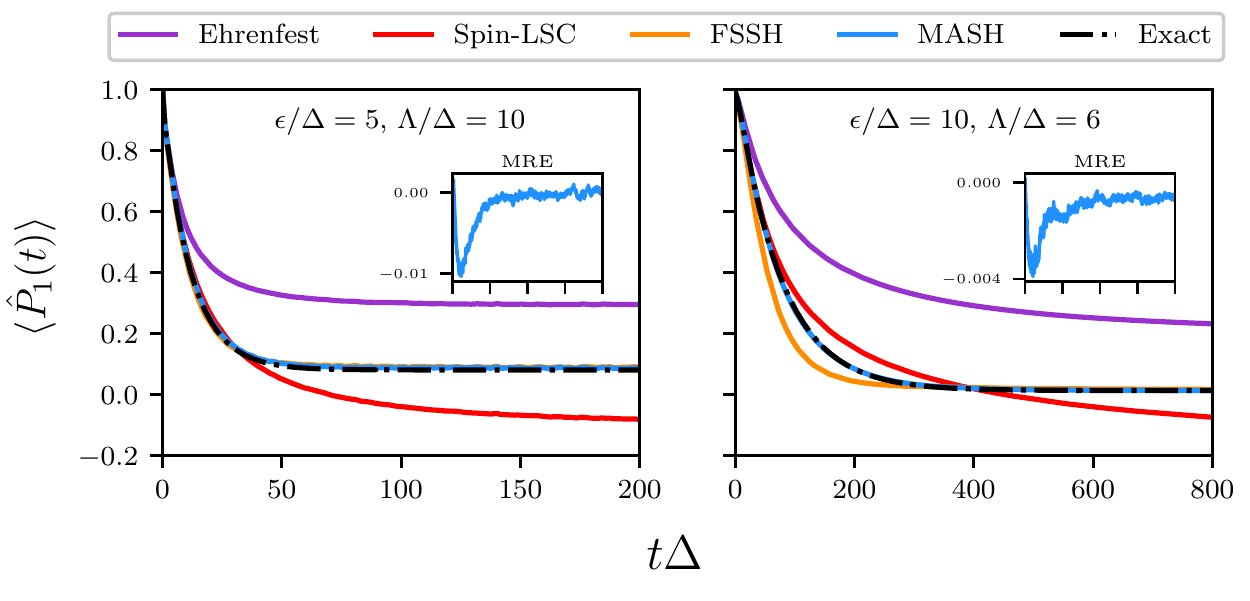}
\caption{The time-dependent diabatic populations for two spin--boson models, as for Fig.~\ref{fig:spin-low}, but with parameters $\beta\Delta=0.25$ and $\omega_{\text{c}}/\Delta=0.2$. The remaining parameters are given in the respective figures.  Note that the system on the left is in the Marcus normal regime, $\epsilon<\Lambda$, whereas the system on the right is in the inverted regime, $\epsilon>\Lambda$.}\label{fig:spin-asym}
\end{figure}
Figure~\ref{fig:spin-asym} shows the diabatic populations for two spin--boson models that undergo incoherent decay. %
It is known that spin-LSC can give rise to negative populations in this regime, as trajectories preferentially occupy the region below the polar circle associated with the lower-energy diabat, $\ket{2}$.\cite{ultrafast} In contrast, both surface-hopping approaches pretty much exactly reproduce the long-time limit of the populations. Because MASH also well describes the population decay timescale in both cases, this suggests that MASH essentially obeys detailed balance in many systems, although mathematically it is not rigorously obeyed (Appendix~\ref{sec:detailed_balance}). Although FSSH captures the correct timescales for the first system in the normal regime of Marcus theory, it produces a significant error for the population decay timescale for the inverted-regime model in the right-hand panel. This problem is absent within MASH, and thus this deficiency is almost certainly another manifestation of the FSSH inconsistency error introduced previously in Sec.~\ref{sec:tully3}. This therefore explains why MASH can already accurately capture the timescales for population decay without the need for decoherence corrections. Because this inconsistency error is also likely to be the reason why decoherence corrections are needed within FSSH to capture Marcus-theory rates,\cite{Landry2011hopping,Landry2012hopping} MASH may remarkably offer a simple approach for calculating accurate nonadiabatic rates without decoherence corrections. We will explore this in future work. %

\subsubsection{Internal Conversion in Pyrazine}
Pyrazine is a classic case for ultrafast relaxation processes involving a conical intersection. It has been well studied with a variety of methods from formally exact quantum mechanics using traditional basis sets\cite{Raab1999,Xin2006,Baiardi2019DMRG} and 
trajectory-based quantum simulations,\cite{BenNun2002AIMS,Burghardt2008,Shalashilin2010} as well as approximate quasiclassical approaches.\cite{Stock2005nonadiabatic,Lang2021GDTWA} After photoexcitation to the S$_{2}$ electronic state, pyrazine relaxes on an ultrafast timescale to the lower-energy S$_{1}$ state. These dynamics can be reasonably well described using vibronic coupling models.

\paragraph{Three-Mode Model}\label{sec:pyrazine3d}
We first consider the following three-mode model,
\begin{subequations}
\label{eq:ham_3d}
\begin{align}
\bar{V}(\bm{q})&=\tfrac{1}{2}\sum_{j=1}^{3}\left[\omega_{j}^{2}q_{j}^{2}+(a_{j}+b_{j})\sqrt{\omega_{j}}q_{j}\right] , \\
\kappa(\bm{q})&=\tfrac{1}{2}[E_{1}-E_{2}]+\tfrac{1}{2}\sum_{j=1}^{3}(a_{j}-b_{j})\sqrt{\omega_{j}}q_{j} , \\
\Delta(\bm{q})&=c_{3}\sqrt{\omega_{3}}q_{3} . 
\end{align}
\end{subequations}
Compared to the completely equivalent Hamiltonian given in Ref.~\onlinecite{Schneider1988} (where all parameters of the model can be found), the additional factors of $\sqrt{\omega_{j}}$ arise because we use mass-weighted coordinates so that $m=1$. The system is initialized in the second diabatic state, $\ket{2}$, with the nuclear phase-space variables sampled from the ground vibrational state [Eq.~(\ref{eq:wavepacket}), with $\bar{q}=\bar{p}=0$ and $\gamma=\omega_{j}$ for each mode, $j$]. While this model is still harmonic like the spin--boson models considered earlier, its dynamics are fundamentally different due to the presence of a conical intersection. Exact results, computed using the discrete-variable representation, were obtained from Ref.~\onlinecite{Muller1997}. 

\begin{figure*}
\includegraphics[scale=1]{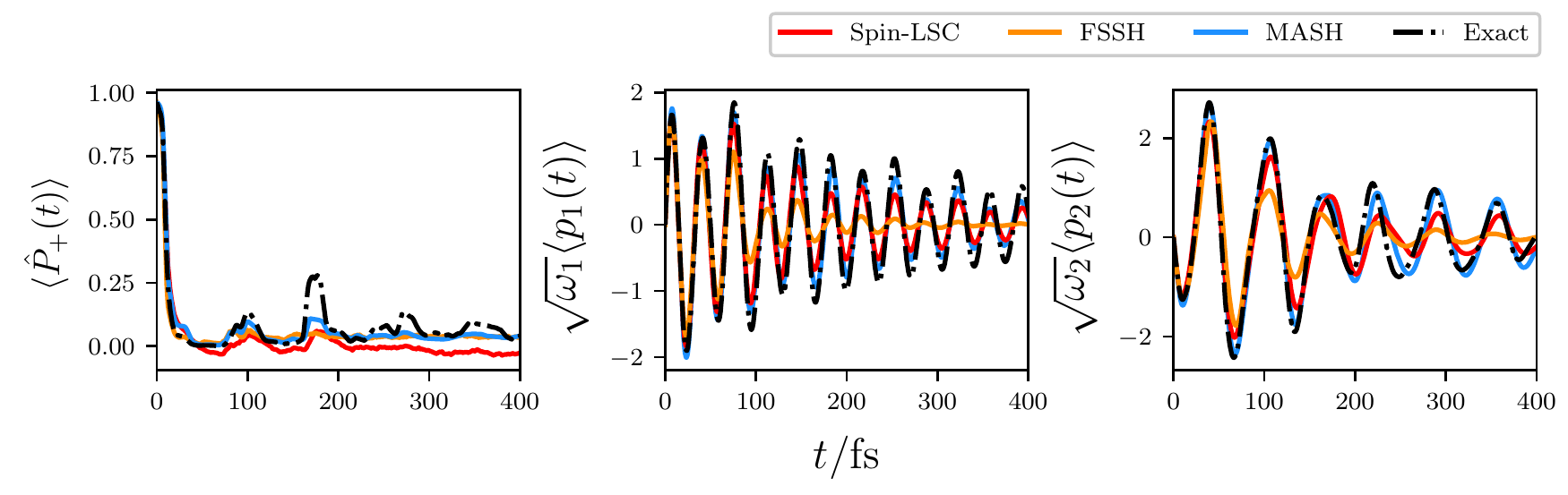}
\caption{The time-dependent population of the upper adiabatic state %
and the nuclear momentum associated with modes 1 and 2 %
for a two-level model of pyrazine containing three nuclear degrees of freedom [Eq.~(\ref{eq:ham_3d})].
The mass-weighting is removed for comparison with Ref.~\onlinecite{Stock2005nonadiabatic}.
}\label{fig:pyrazine3d}
\end{figure*}
The left panel of Fig.~\ref{fig:pyrazine3d} presents the population of the upper adiabat. Spin-LSC again gives rise to negative populations in the long-time limit, however it does so less severely than MMST.\cite{Stock2005nonadiabatic,JohanPhD} 
In contrast, the populations from both surface-hopping approaches always remain positive and although they cannot accurately capture the recurrences, they match the exact result well. %

The middle and right-hand panels give the expectation value of the nuclear momentum for two of the modes.  It can be seen that FSSH overdamps these oscillations. This was also noticed in previous work,\cite{Muller1997} in which it was found that better agreement could be found when momentum reversals were not applied for frustrated hops. It was suggested that momentum reversals should hence not be included in the FSSH method. However, we find that MASH (which does include momentum reversals for frustrated hops) accurately reproduces the magnitude of these oscillations. This suggests that the overdamped oscillations obtained for FSSH instead arise due to an error that is absent in MASH, such as the inconsistency error that only occurs in FSSH\@.
The reason that the FSSH results appear to improve when neglecting momentum reversals must therefore be due to a fortuitous cancellation of errors.
The fact that momentum reversals for frustrated hops arise naturally within MASH from its derivation from the QCLE (Appendix~\ref{sec:QCLE}) %
implies that Tully's original suggestion\cite{Hammes-Schiffer1994} to employ momentum reversals in FSSH is the right approach.
Instead of suggesting alternatives to momentum reversals, MASH has fixed the inconsistency of FSSH, which ultimately leads to its improved results.
\paragraph{24-Mode Model}\label{sec:pyrazine24d}
Here we consider an extended model of pyrazine, which not only contains more vibrational degrees of freedom, but also contains bilinear couplings that give rise to effects beyond the model systems treated above.  In particular, the frequencies of the harmonic oscillators are different in the two electronic states. The Hamiltonian for this model is given by
\begin{subequations}
\label{eq:ham_24d}
\begin{align}
\begin{split}
\bar{V}(\bm{q})=&\tfrac{1}{2}\sum_{j=1}^{24}\Bigg[\omega_{j}^{2}q_{j}^{2}+(a_{j}+b_{j})\sqrt{\omega_{j}}q_{j} \\
&\qquad\qquad\quad+\sum_{k=1}^{24}(a_{jk}+b_{jk})\sqrt{\omega_{j}\omega_{k}}q_{j}q_{k}\Bigg] ,
\end{split} \\
\begin{split}
\kappa(\bm{q})=&\tfrac{1}{2}[E_{1}-E_{2}]+\tfrac{1}{2}\sum_{j=1}^{24}\Bigg[(a_{j}-b_{j})\sqrt{\omega_{j}}q_{j} \\
&\qquad\qquad\,\,\,\,+\sum_{k=1}^{24}(a_{jk}-b_{jk})\sqrt{\omega_{j}\omega_{k}}q_{j}q_{k}\Bigg] , 
\end{split} \\
\Delta(\bm{q})=&\sum_{j=1}^{24}\left[c_{j}\sqrt{\omega_{j}}q_{j}+\sum_{k=1}^{24}c_{jk}\sqrt{\omega_{j}\omega_{k}}q_{j}q_{k}\right] , 
\end{align}
\end{subequations}
where %
the parameters can be found in Ref.~\onlinecite{Raab1999}. Additional factors of $\sqrt{\omega_{j}}$ have been included in our expression, because we choose to use mass-weighted coordinates for which $m=1$. The initial conditions for the dynamics are the same as for the three-mode model introduced earlier in Sec.~\ref{sec:pyrazine3d}.

\begin{figure}
\includegraphics[scale=1]{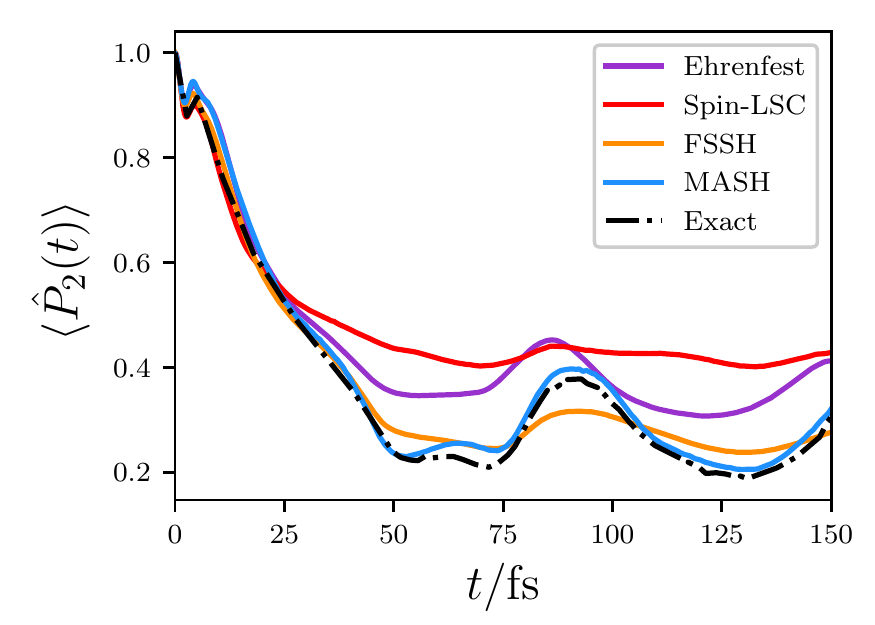}
\caption{The time-dependent population of the second diabatic state, $P_{2}(t)$, for a two-level model of pyrazine containing 24 nuclear degrees of freedom [Eq.~(\ref{eq:ham_24d})]. %
}\label{fig:pyrazine24d}
\end{figure}
Figure~\ref{fig:pyrazine24d} presents the computed diabatic populations. The numerically exact results, computed using the multi-configuration time-dependent Hartree (MCTDH) method, were obtained from Ref.~\onlinecite{Raab1999_2}. Spin-LSC produces the largest error in the dynamics for this system, because some trajectories acquire negative populations, leading to unphysical forces that correspond to propagating the nuclei on inverted potentials. While this is not such a big problem for the linearly-coupled harmonic bath models considered previously, the presence of bilinear couplings in this model mean that the inverted potentials can become unbounded and the trajectories can accelerate off to infinity. Unusually, Ehrenfest even outperforms spin-LSC in this case, because the former cannot have negative populations or propagate on inverted potentials. While both FSSH and MASH are able to on average correctly reproduce the long-time limit of the populations, MASH is much better at describing the magnitude of the oscillations, producing results very close to the benchmark for the entire dynamics. Note that the MASH simulation is no more expensive than FSSH and because independent classical-trajectory methods like MASH are significantly less computationally expensive than quantum wavepacket propagation approaches like MCTDH, the MASH dynamics can easily be propagated for larger and more complex systems containing anharmonic potential-energy surfaces and even used for on-the-fly \emph{ab initio} simulations.

\section{Discussion} %

It is interesting to ask why MASH is able to make use of deterministic trajectories, whereas stochastic trajectories are required within FSSH\@.
The FSSH method is essentially an extension of the Ehrenfest formalism, in that it uses the Bloch sphere and initializes the electronic variables at a single point in order to represent a pure state.
Thus, in the simplest case of scattering boundary conditions, all trajectories start with the same initial conditions and, in order to have a distribution of product states, it is necessary for a random element to enter into the dynamics.
In contrast, the MASH method is based on the mapping formalism, which initializes the mapping variables in a distribution smeared out over the spin-sphere.
Each trajectory has a different initial condition and therefore different final states can arise even though the equations of motion are fully deterministic.
Thus, it is fundamentally because MASH is built on the mapping approach rather than Ehrenfest that leads to the key differences with FSSH.

It is also worthwhile to compare and contrast MASH with the SQC method.\cite{Miller2016Faraday}
Both approaches are based on mapping the electronic states to a phase space and using windowing schemes to determine the electronic state of an individual trajectory when measuring an observable.
In SQC, it was observed that significant improvements in the results could be obtained (especially in the weak-$\Delta$ limit) by ensuring that the windows touched.\cite{Cotton2016SQC}
The MASH windows also touch and it is clear from the derivation in Appendix~\ref{sec:Rabi} that this is necessary in order to obtain the correct dynamics within the classical path approximation.
There are, however, two key differences.
First, the SQC method is based on MMST mapping, which has a large phase space including both physical and unphysical regions.\cite{Mueller1998mapping}
The SQC windows therefore do not fill the space and if trajectories start or finish outside the windows, they are discarded.
Because MASH is based on spin mapping, which has a reduced phase space corresponding only to physical states, it is possible to construct windows which fill the space.
It is not therefore necessary to discard any trajectories.
It is, however, necessary to introduce a positive-definite weighting factor into the MASH correlation function in order to recover the correct electronic dynamics within the classical path approximation.
SQC effectively does this by tailoring the shape of the triangular windows.\cite{thermalization}
Second, the equations of motion in the SQC method use the same mean-field forces as in linearized mapping methods.
MASH, on the other hand, also `windows' its forces and thus can capture wavepacket branching and does not suffer from inverted potentials.
It is not obvious how one would generalize SQC in this way as trajectories may pass outside the windows (as long as they return at a later time to contribute to the observable) and it is not clear how the force at these intermediate times should be defined.
Overall, therefore, it is only because we employ the spin-mapping formalism rather than MMST that it is possible to construct the MASH method.

The success of surface-hopping methods relies on an assumption that the true quantum dynamics is well approximated by classical trajectories moving on \emph{adiabatic} potential-energy surfaces. 
Tully often points out that there are circumstances in which the Ehrenfest method outperforms FSSH, especially in cases where there are many states close in energy, such as near metal surfaces.\cite{Tully1998MQC}
In the same way, there is no guarantee that MASH will be superior to spin-LSC, as both approaches are formally derived as short-time approximations to the QCLE\@.
In fact, in this work we have presented a few examples where the mean-field force of spin-LSC is more appropriate, such as for the strongly coherent dynamics in Tully's dual avoided crossing model (II) and also in some, but not all, parameter regimes of the spin-boson model.
The cases where MASH is expected to exhibit the most dramatic improvements over spin-LSC are in the strongly anharmonic systems encountered in photochemistry, which will be the focus of future work.

In our derivation of MASH, we have found that the momentum should be reversed at every frustrated hop and that this is necessary in order to retain the rigorous connection to the QCLE and to enable systematic improvements using the quantum-jump procedure.
Although this is what Tully originally suggested,\cite{Hammes-Schiffer1994}
Jasper and Truhlar found that results could be improved by only reversing the momentum in certain situations.\cite{Jasper2003}
They noted that reversing the momentum is beneficial for predicting nonadiabatic transition probabilities and branching ratios, %
but that it is better to not reverse the momentum in order to obtain vibrational and rotational quantum numbers (similar to M\"uller and Stock's finding).\cite{Muller1997}
Their so-called `$\nabla V$' prescription is to only reverse the momentum if the momentum and the force of the upper state (both projected in the direction of the nonadiabatic coupling vector) have opposite signs.
In our study of the three-mode pyrazine model, we show that MASH can accurately recover the vibrational dynamics even though we do apply momentum reversals for every frustrated hop.
It therefore appears that the \textit{ad hoc} `$\nabla V$' prescription is not required in MASH\@.
This is because it solves the inconsistency problem of FSSH and, thus, does not require a work-around.

It is interesting to note that our finding that the momentum rescalings are rigorously derivable is in complete contrast to the conclusions of Martens, who proposed that they should not be employed at all (even for allowed hops).
Martens derived a coupled-trajectory method \cite{Martens2016} and a more practical, approximate independent-trajectory method.\cite{Martens2019,Martens2020}
Martens' methods and MASH are both founded on the QCLE and both are modified versions of FSSH, %
but in completely different ways.
For instance, his trajectories are stochastic and employ a quantum force instead of momentum rescaling, meaning that energy is conserved only on average, but not at an individual trajectory level.
It will be interesting to compare the advantages and disadvantages of the two methods in future work.

\section{Conclusions}
In this paper we have obtained a rigorous mapping approach to surface hopping (MASH) that offers the best of both worlds between FSSH and quasiclassical mapping methods. By using forces from a single adiabatic state, MASH has no problem with simulating strongly anharmonic systems, which can cause the mean-field forces of spin-LSC and other mapping methods to behave unphysically. Additionally, MASH alleviates many of the philosophical and practical problems of FSSH\@. In particular, the dynamics are deterministic, which guarantees that the active surface and electronic state always remain consistent. This means that MASH can be used to obtain accurate predictions in many cases without the need for decoherence corrections. We have demonstrated this by applying MASH to a range of popular nonadiabatic model systems, where MASH was found to be the most reliable classical-trajectory approach overall.

For the cases where MASH is unable to capture the exact dynamics, a rigorous quantum-jump procedure can be applied to systematically improve the accuracy of the results to those of the QCLE\@. This jump procedure is able to describe quantum interference phenomena, which cannot be included by standard FSSH decoherence schemes. For example, the MASH jump procedure was able to systematically improve the wavepacket dynamics of Tully's model II towards the exact result, while commonly used FSSH decoherence corrections are seen to have no effect.\cite{Subotnik2011AFSSH} Additionally, unlike FSSH decoherence corrections, jumps do not formally degrade the quality of the results, even if they are applied at suboptimal times or more frequently than necessary (although they do require more trajectories to achieve numerical convergence). For the situations where the MASH error is solely due to an incorrect description of decoherence, such as for Tully's model III, an approximate decoherence correction can be derived from the rigorous jump procedure, which only had to be applied once to fully correct the dynamics. This decoherence correction in general requires far fewer trajectories to reach numerical convergence than for the quantum jumps. However, as for FSSH decoherence corrections, the MASH version can also degrade the quality of results if it is applied at the wrong time or too often. It should be possible to combine our new decoherence scheme with those already used within FSSH (such as augmented FSSH),\cite{Subotnik2011AFSSH} in order to efficiently determine the appropriate times at which the decoherence correction should be applied during simulations of more complex systems.

One useful observation is that MASH approximately obeys detailed balance in condensed-phase problems, such as for spin--boson models in the classical-nuclear limit. This can be seen in our simulations from the fact that MASH can very accurately predict the long-time limit of correlation functions for models in this regime. We have, however, shown that the approach does not rigorously obey detailed balance in general, due to non-conservation of the MASH weighting factor, which formally violates microscopic reversibility. Using this result, a microscopic reversibility error (MRE) can be constructed that is by definition zero in the regimes where the MASH dynamics obeys microscopic reversibility. We have computed the MRE for each model considered in this paper and for the majority of cases it is found to remain very close to zero, in keeping with our original observation. \footnote{We did not show the MRE for Tully's model I and the two pyrazine models, because it always remained very small throughout the dynamics, never exceeding a value of 0.035.} In the cases where it deviated significantly from zero, we observed that there was a corresponding error in the populations. While giving an indication of when the obtained dynamics can be trusted, the MRE may also prove useful for indicating when jumps or approximate decoherence corrections should be applied in a simulation.

In order to ensure full convergence of our results and to capture the subtle differences between the various methods, we used $10^{6}$ trajectories for all quasiclassical methods in this work, including the case in which the MASH decoherence correction was applied to Tully's model III\@.
In practice, however, it is often sufficient to use only about 1000 trajectories in FSSH to obtain qualitatively correct results with only a few small wiggles.\cite{Subotnik2011AFSSH}
We observed that the convergence behaviour of MASH is similar to that of FSSH, so that in practice, far fewer trajectories than we have actually used are needed in the majority of cases to qualitatively describe the correct dynamics. 
Of course more trajectories are in general required for the full quantum-jump procedure; for instance, $8\times10^{7}$ were used for the $n_{\text{jump}}=4$ calculations for Tully's model II (Fig.~\ref{fig:tully2_packet}).  Further approximations will be needed to bring this number down to a manageable size in practical applications.

The similar computational cost of FSSH and MASH is not surprising, %
as the algorithms are strongly related.
Because of this,
many of the tricks developed to make FSSH faster and more numerically stable\cite{Granucci2001,Barbatti2014newtonX,Mai2018SHARC} can be incorporated directly into MASH\@. 
This includes machine-learning methodology for predicting the potentials and nonadiabatic couplings from \textit{ab initio} training data.\cite{Westermayr2019ML,Menger2020PySurf,MLNACs}
The main difference between the two approaches is that instead of using random numbers to determine when to hop in FSSH,  random numbers are incorporated into the initial sampling of the spin-coordinates in MASH.  In both cases, we found that a similar number of trajectories were required to reach convergence. Although not a crucial point, an area of future work could be to investigate more efficient sampling schemes for the initial spin-coordinates, in order to further increase the efficiency of MASH\@. This would be particularly advantageous for one-dimensional scattering problems, where the only sampling involved is over the spin-sphere, which for example could be performed even more efficiently with quadrature.

In this paper, MASH has only been derived for two-level systems. Therefore the most pressing aspect of future work is to generalize the underlying theory to multi-state problems as has been done for spin-LSC\cite{multispin} and SQC.\cite{Cotton2019FMO}
We assume that this will require windows that (unlike for SQC) fill the full mapping space and weighting factors for the correlation functions that allow the correct description of the dynamics within the classical path approximation. Additionally, it would also be desirable to extend the MASH framework to calculate other dynamical quantities of interest, such as multi-time correlation functions. In other work, we have already shown how partially-linearized spin-mapping methods can be used to compute nonlinear spectra.\cite{nonlinear} It would therefore be useful to derive a partially-linearized version of MASH. %
Finally we have also found that MASH is better than FSSH at describing the timescales of population decay, which suggests that MASH could be both a useful method for calculating nonadiabatic rates and a good starting point for an improved nonadiabatic ring-polymer theory.\cite{mapping,Ananth2013MVRPMD,Chowdhury2017CSRPMD,Tao2018isomorphic} %

\begin{acknowledgments}
The authors acknowledge the support from the Swiss National Science Foundation through the NCCR MUST (Molecular Ultrafast Science and Technology) Network. We thank Graziano Amati, Johan Runeson, Joseph Lawrence and Basile Curchod for helpful discussions and comments. In particular, we would also like to thank Johan Runeson for inspiration for the acronym, MASH. 
\end{acknowledgments}

\section*{Author Declarations}
\subsection*{Conflict of Interest}
The authors have no conflicts to declare.
\subsection*{Author Contributions}
\textbf{Jonathan R. Mannouch}: Conceptualization (lead); Formal analysis (lead); Investigation (lead); Methodology (lead); Software (lead); Writing - original draft (lead); Writing - review \& editing (equal). %
\textbf{Jeremy O. Richardson}: Conceptualization (supporting); Formal analysis (supporting); Methodology (supporting); Writing - review \& editing (equal). %

\section*{Data Availability}
The data that supports the findings of this study are available within the article.

\appendix

\section{Pauli Matrices in the Adiabatic Basis}\label{sec:transf}
The Pauli matrices, along with the identity operator, provide a convenient basis for representing any Hermitian operator for a two-state system. In this paper, we refer to the adiabatic Pauli matrices, $\hat{\bm{\sigma}}(\bm{q})$, as those constructed in the adiabatic basis, so that they diagonalize the electronic contribution to the Hamiltonian. They have the usual commutation relations and depend on the nuclear coordinates, $\bm{q}$, due to their connection to the adiabatic basis. For a state-dependent potential defined in the diabatic basis [Eq.~(\ref{eq:ham_diab})], the adiabatic Pauli matrices are given by
\begin{subequations}
\label{eq:transform}
\begin{align}
\hat{\sigma}_{x}(\bm{q})&=\frac{\kappa(\bm{q})\hat{\sigma}^{\text{diab}}_{x}-\Delta(\bm{q})\hat{\sigma}^{\text{diab}}_{z}}{\sqrt{\kappa(\bm{q})^{2}+\Delta(\bm{q})^{2}}} , \\
\hat{\sigma}_{y}(\bm{q})&=\hat{\sigma}^{\text{diab}}_{y} \\
\hat{\sigma}_{z}(\bm{q})&=\frac{\Delta(\bm{q})\hat{\sigma}^{\text{diab}}_{x}+\kappa(\bm{q})\hat{\sigma}^{\text{diab}}_{z}}{\sqrt{\kappa(\bm{q})^{2}+\Delta(\bm{q})^{2}}}  ,
\end{align}
\end{subequations}
where $\hat{\bm{\sigma}}^{\text{diab}}$ are the usual Pauli matrices in the diabatic basis, which do not depend on the nuclear coordinates, $\bm{q}$. Equivalently the reverse transformations associated with Eqs.~(\ref{eq:transform}) are given by
\begin{subequations}
\label{eq:transform_rev}
\begin{align}
\hat{\sigma}^{\text{diab}}_{x}&=\frac{\kappa(\bm{q})\hat{\sigma}_{x}(\bm{q})+\Delta(\bm{q})\hat{\sigma}_{z}(\bm{q})}{\sqrt{\kappa(\bm{q})^{2}+\Delta(\bm{q})^{2}}} , \\
\hat{\sigma}^{\text{diab}}_{y}&=\hat{\sigma}_{y}(\bm{q}) , \\
\hat{\sigma}^{\text{diab}}_{z}&=\frac{-\Delta(\bm{q})\hat{\sigma}_{x}(\bm{q})+\kappa(\bm{q})\hat{\sigma}_{z}(\bm{q})}{\sqrt{\kappa(\bm{q})^{2}+\Delta(\bm{q})^{2}}} \label{eq:transform_z}.
\end{align}
\end{subequations}
By substitution of Eqs.~(\ref{eq:transform_rev}) into Eq.~(\ref{eq:ham_diab}), it can be shown that the adiabatic Pauli matrices diagonalize the state-dependent potential, as required, giving adiabatic potential-energy surfaces, $V_{\pm}(\bm{q})=\bar{V}(\bm{q})\pm V_{z}(\bm{q})$, with
\begin{equation}
V_{z}(\bm{q})=\sqrt{\kappa(\bm{q})^{2}+\Delta(\bm{q})^{2}} .    
\end{equation}

In order to obtain an expression for the nonadiabatic coupling vectors in terms of the diabatic Hamiltonian parameters, the equations of motion for the adiabatic Pauli matrices are derived by taking time-derivatives of both sides of Eqs.~(\ref{eq:transform}) and then the chain rule is used to evaluate the time-derivatives of the diabatic Hamiltonian parameters, i.e., $\dot{\kappa}(\bm{q})=\sum_{j}\frac{\partial\kappa(\bm{q})}{\partial q_{j}}\frac{p_{j}}{m}$. Additionally, the following equations of motion for the diabatic Pauli matrices are used (obtained from $\rd\hat{\sigma}^{\text{diab}}_{k}/\rd t=\iu[\hat{V}(\bm{q}),\hat{\sigma}^{\text{diab}}_{k}]$)
\begin{subequations}
\label{eq:mapping_eom_diab}
\begin{align}
\frac{\rd}{\rd t}\hat{\sigma}^{\text{diab}}_{x}&=-2\kappa(\bm{q})\hat{\sigma}^{\text{diab}}_{y} , \\
\frac{\rd}{\rd t}\hat{\sigma}^{\text{diab}}_{y}&=2\kappa(\bm{q})\hat{\sigma}^{\text{diab}}_{x}-2\Delta(\bm{q})\hat{\sigma}^{\text{diab}}_{z} , \\
\frac{\rd}{\rd t}\hat{\sigma}^{\text{diab}}_{z}&=2\Delta(\bm{q})\hat{\sigma}^{\text{diab}}_{y} ,
\end{align}
\end{subequations}
which lead to the same equations of motion as Eq.~(\ref{eq:mapping_eom}), with the nonadiabatic coupling vectors expressed in terms of the diabatic parameters as
\begin{equation}
d_{j}(\bm{q})=\frac{\Delta(\bm{q})\frac{\partial\kappa(\bm{q})}{\partial q_{j}} - \kappa(\bm{q})\frac{\partial\Delta(\bm{q})}{\partial q_{j}}}{2\left(\kappa(\bm{q})^{2}+\Delta(\bm{q})^{2}\right)} .
\end{equation}

In order to initialize Ehrenfest, spin-LSC and FSSH trajectories, Eqs.~(\ref{eq:transform_rev}) can also be used to determine the spin-vector, $\bm{z}_{\alpha}$, on the Bloch sphere that corresponds to the following initial diabatic state 
\begin{equation}
\ket{\alpha}\!\bra{\alpha}=\frac{1}{2}\left[\hat{\mathcal{I}}+\bm{z}_{\alpha}\cdot\hat{\bm{\sigma}}(\bm{q})\right] .  
\end{equation}
For example, using $\hat{P}_{1}=\half[\hat{\mathcal{I}}+\hat{\sigma}_{z}^{\text{diab}}]$ along with the expression given in Eq.~(\ref{eq:transform_z}), we find that the spin-vector associated with $\hat{P}_{1}$ is given by
\begin{equation}
\label{eq:z_diab}
\bm{z}_{\alpha}=\Bigg\{-\frac{\Delta(\bm{q})}{\sqrt{\kappa(\bm{q})^{2}+\Delta(\bm{q})^{2}}},0,\frac{\kappa(\bm{q})}{\sqrt{\kappa(\bm{q})^{2}+\Delta(\bm{q})^{2}}}\Bigg\} . 
\end{equation}
\section{The Classical Path Approximation}\label{sec:Rabi}
For any mapping approach, we would like to exactly describe the bare electronic dynamics within the classical path approximation (CPA), where the nuclear path is pre-specified [$q_{j}(t)$ and $p_{j}(t)=m\dot{q}_{j}(t)$]. If this is correctly described, it guarantees that the approach will be exact in the limit that the state-dependent nuclear force operator can be neglected, either because $\bar{F}(\bm{q})$ dominates or because the initial momentum is so large. 

Under this approximation, the electronic degrees of freedom are evolved according to a time-dependent pure electronic Hamiltonian.
For such a case, Eq.~(\ref{eq:mapping_eom}) shows that the time-dependent spin-coordinates depend on their initial values according to
\begin{equation}
\label{eq:mapping_Rabi}
S_{k}(t)=\tfrac{1}{2}\big[C^{\text{CPA}}_{\sigma_{x}\sigma_{k}}(t)S_{x}+C^{\text{CPA}}_{\sigma_{y}\sigma_{k}}(t)S_{y}+C^{\text{CPA}}_{\sigma_{z}\sigma_{k}}(t)S_{z}\big],
\end{equation}
where $C^{\text{CPA}}_{\sigma_{\ell}\sigma_{k}}(t)$ for $\ell,k\in\{x,y,z\}$ is the exact quantum correlation function for this pre-specified nuclear path.

It has been shown that the spin-LSC method (as for many other mapping approaches) exactly reproduces correlation functions for a bare electronic system.\cite{spinmap} In MASH, however, we need to reconsider the question, because although the spin dynamics is the same, the measurement procedure is different. In particular we want there to be internal consistency between how the nuclei and observables see the electronic state of the system. This requires that the mapping-variable representations of the electronic operators must be given by Eqs.~(\ref{eq:density_MASH}). We now consider how to construct the various possible correlation functions with these operator representations in a way that the dynamics within the classical path approximation are correctly described. 
\subsection{The Coherence--Coherence Correlation Functions}
The coherence--coherence correlation function is the trivial case, because the dynamics of the spin-coordinates [Eq.~(\ref{eq:mapping_eom})] are designed by construction to reproduce the bare electronic dynamics for observable operators directly represented by the mapping variables themselves. Using the fact that $\int\rd\bm{S}\,S_{\ell}S_{k}=\tfrac{2}{3}\delta_{\ell k}$, where $\delta_{\ell k}$ is the Kronecker delta, it can be shown from Eq.~(\ref{eq:mapping_Rabi}) that
\begin{equation}
C^{\text{CPA}}_{\sigma_{\ell}\sigma_{k}}(t)=3\int\rd\bm{S}\,S_{\ell}S_{k}(t) , \quad\text{for } \ell,k\in\{x,y\},
\end{equation}
where the integral over spin-coordinates is defined in Eq.~(\ref{eq:spin_int}). This is the same expression for the correlation function used in full-sphere spin-LSC\cite{spinmap} (with $r_{\text{s}}^2=3$) and determines the value $\mathcal{W_{\text{CC}}}(\bm{S})=3$, as given by the first entry on the right-hand side of Eq.~(\ref{eq:MASH_weight}).
\subsection{The Population--Coherence Correlation Functions}\label{sec:pop_coh}
We next consider the simple case of starting in an adiabatic population and then measuring a coherence at time $t$. In the case of purely electronic dynamics under a time-dependent Hamiltonian, we find that $C^{\text{CPA}}_{P_{\pm}\sigma_{k}}(t)=\pm\tfrac{1}{2}C^{\text{CPA}}_{\sigma_{z}\sigma_{k}}(t)$ for $k\in\{x,y\}$, as can be shown from the fact that $P_{\pm}(\bm{q})=\tfrac{1}{2}[\hat{\mathcal{I}}\pm\hat{\sigma}_{z}(\bm{q})]$ and $C^{\text{CPA}}_{\mathcal{I}\sigma_{k}}(t)=\int\rd\bm{S}\,S_{k}(t)=0$ by symmetry. Finally, using $\int\rd\bm{S}\,h(\pm S_{z})S_{k}=\pm\tfrac{1}{2}\delta_{zk}$, we find that
\begin{equation}
\label{eq:rabi_mix}
C^{\text{CPA}}_{P_{\pm}\sigma_{k}}(t)=2\int\rd\bm{S}\,h(\pm S_{z})S_{k}(t), \quad\text{where } k\in\{x,y\} ,
\end{equation}
which determines the value of $\mathcal{W_{\text{PC}}}(\bm{S})=2$ that interestingly corresponds to the MMST radius, $r_{s}=2$.

A correlation function that is initialized in a coherence and measures an adiabatic population at time $t$ is also guaranteed to reproduce the bare electronic dynamics, given that this is true for the opposite case too. This is because for a pure electronic system, the spin-mapping equations of motion are time-translationally invariant and preserve phase-space volume, such that
\begin{equation}
\int\rd\bm{S}\,S_{k}h(\pm S_{z}(t))=\int\rd\bm{S}\,S_{k}(-t)h(\pm S_{z}) ,
\end{equation}
which is connected to $C^{\text{CPA}}_{P_{\pm}\sigma_{k}}(t)$ by time-reversal symmetry. This integral can be performed in the same way as for Eq.~(\ref{eq:rabi_mix}), giving the required result that $\mathcal{W_{\text{CP}}}(\bm{S})=2$.
\subsection{The Population--Population Correlation Functions}
In contrast to the previous two cases, the population--population correlation functions require a careful treatment. Here we will consider the $C^{\text{CPA}}_{P_{+}P_{+}}(t)$ correlation, as the same proof can be easily used to obtain the other three population--population correlation functions. Using $P_{+}(\bm{q})=\tfrac{1}{2}[\hat{\mathcal{I}}+\hat{\sigma}_{z}(\bm{q})]$ as before, we find that $C^{\text{CPA}}_{P_{+}P_{+}}(t)=\tfrac{1}{2}[1+\tfrac{1}{2}C^{\text{CPA}}_{\sigma_{z}\sigma_{z}}(t)]$, because $C^{\text{CPA}}_{\mathcal{I}\sigma_{z}}(t)=C^{\text{CPA}}_{\sigma_{z}\mathcal{I}}(t)=0$ for dynamics associated with a pure electronic system. 

We start by showing that the naive MASH form for this population--population correlation does not reproduce the correct bare electronic dynamics. Namely
\begin{equation}
\label{eq:zz_wrong}
C^{\text{CPA}}_{P_{+}P_{+}}(t)\neq\int\rd\bm{S}\,h(S_{z})h(S_{z}(t)) .
\end{equation}
To explain this, we start by introducing new coordinates
\begin{subequations}
\begin{align}
S'_{x}&=-\frac{1}{2}\frac{C^{\text{CPA}}_{\sigma_{x}\sigma_{z}}(t)S_{x}+C^{\text{CPA}}_{\sigma_{y}\sigma_{z}}(t)S_{y}}{\sqrt{[\tfrac{1}{2}C^{\text{CPA}}_{\sigma_{x}\sigma_{z}}(t)]^{2}+[\tfrac{1}{2}C^{\text{CPA}}_{\sigma_{y}\sigma_{z}}(t)]^{2}}} , \\
S'_{y}&=\frac{1}{2}\frac{C^{\text{CPA}}_{\sigma_{y}\sigma_{z}}(t)S_{x}-C^{\text{CPA}}_{\sigma_{x}\sigma_{z}}(t)S_{y}}{\sqrt{[\tfrac{1}{2}C^{\text{CPA}}_{\sigma_{x}\sigma_{z}}(t)]^{2}+[\tfrac{1}{2}C^{\text{CPA}}_{\sigma_{y}\sigma_{z}}(t)]^{2}}} ,
\end{align}
\end{subequations}
so that $S_{z}(t)$ originally defined in Eq.~(\ref{eq:mapping_Rabi}) has no dependence on $S'_{y}$ and can be written as
\begin{equation}
\label{eq:mapping_Rabi2}
S_{z}(t)=-\sqrt{1-[\tfrac{1}{2}C^{\text{CPA}}_{\sigma_{z}\sigma_{z}}(t)]^{2}}S'_{x}+\tfrac{1}{2}C^{\text{CPA}}_{\sigma_{z}\sigma_{z}}(t)S_{z} ,
\end{equation}
where we have additionally used  $[\tfrac{1}{2}C^{\text{CPA}}_{\sigma_{x'}\sigma_{z}}(t)]^2+[\tfrac{1}{2}C^{\text{CPA}}_{\sigma_{y'}\sigma_{z}}(t)]^2+[\tfrac{1}{2}C^{\text{CPA}}_{\sigma_{z}\sigma_{z}}(t)]^2=1$. The phase-space points that give a non-zero contribution to the right-hand side of Eq.~(\ref{eq:zz_wrong}) correspond to the spin-coordinates for which both Heaviside step functions are simultaneously non-zero. This corresponds to all values of the spin-coordinates which are simultaneously in the upper hemisphere (for the first step function to be non-zero) and above the plane $\tfrac{1}{2}C^{\text{CPA}}_{\sigma_{z}\sigma_{z}}(t)S_{z}=\sqrt{1-[\tfrac{1}{2}C^{\text{CPA}}_{\sigma_{z}\sigma_{z}}(t)]^{2}}S'_{x}$ (for the second step function to be non-zero). Defining these spin-variables in polar coordinates as
\begin{subequations}
\begin{align}
S'_{x}&=\sin\theta'\cos\phi' , \\
S'_{y}&=\cos\theta' , \\
S_{z}&=\sin\theta'\sin\phi' ,
\end{align}
\end{subequations}
this corresponds to $\gamma(t)=\cos^{-1}\left[\tfrac{1}{2}C^{\text{CPA}}_{\sigma_{z}\sigma_{z}}(t)\right]\leq\phi'\leq\pi$. Using this information, the integral on the right-hand side of Eq.~(\ref{eq:zz_wrong}) can be evaluated analytically to give
\begin{equation}
\label{eq:pop_pop}
\begin{split}
\int\rd\bm{S}\,&h(S_{z})h(S_{z}(t)) \\
&=\frac{1}{2\pi}\int_{0}^{\pi}\rd\theta'\,\sin\theta'\int_{\gamma(t)}^{\pi}\rd\phi' \\
&=1-\frac{1}{\pi}\gamma(t) ,
\end{split}
\end{equation}
which is not the correct expression for the population--population correlation function for a bare electronic system.

A simple way of fixing this problem is to introduce a weighting factor into the integral. For example, if we use $\mathcal{W}_{\text{PP}}(\bm{S})=2|S_{z}|$, we find
\begin{equation}
\label{eq:pop_pop_weight}
\begin{split}
2\int\rd\bm{S}\,&|S_{z}|h(S_{z})h(S_{z}(t)) \\
&=\frac{1}{\pi}\int_{0}^{\pi}\rd\theta'\,\sin^{2}\theta'\int_{\gamma(t)}^{\pi}\rd\phi'\sin(\phi') \\
&=\frac{1}{2}\left[1+\tfrac{1}{2}C^{\text{CPA}}_{\sigma_{z}\sigma_{z}}(t)\right] ,
\end{split}
\end{equation}
which now correctly reproduces the exact quantum-mechanical result for the bare electronic dynamics. Note that for a correlation involving an initial $\hat{\sigma}_{z}(\bm{q})$ operator, an alternative way of thinking about the weighting is that it can be contracted with the operator representation to give $|S_z|[h(S_z)-h(-S_{z})]= S_{z}$, making one operator equivalent to the spin-LSC case and leading to a situation akin to Appendix~\ref{sec:pop_coh}. By this reasoning, $\mathcal{W}_{\text{PP}}(\bm{S})=2|S_{z}(t)|$ would also be a valid weighting factor and it can be shown that this also reproduces both the bare electronic dynamics and the exact dynamics in the Born--Oppenheimer limit. In certain situations it may also give more accurate results, because $2|S_{z}(t)|$ is expected to better represent the current state of a trajectory after going through the first avoided crossing and before undergoing subsequent transitions. However, because this weight is not guaranteed to enforce that the sum of time-evolved populations is equal to one, we will use $\mathcal{W}_{\text{PP}}(\bm{S})=2|S_{z}|$ throughout this paper.

\begin{figure}
\includegraphics[scale=1]{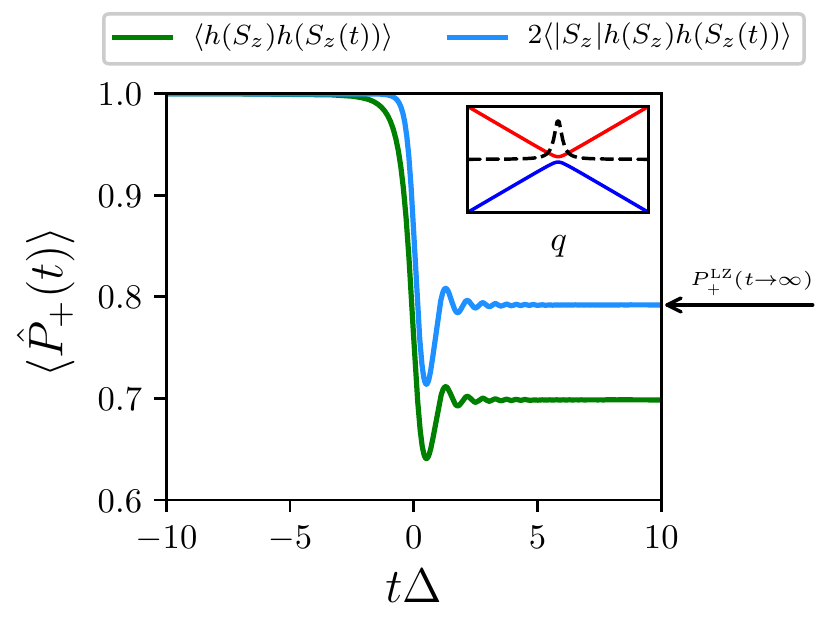}
\caption{The population dynamics for a Landau--Zener model with $p_{\text{const}}/\Delta=2$. This figure demonstrates that the weighting factor, $\mathcal{W}_{\text{PP}}(\bm{S})=2|S_{z}|$, is necessary within MASH, because only the blue line reproduces the correct adiabatic population transfer of the exact quantum-mechanical result (black arrow). Additionally the inset plots the upper and lower adiabatic surfaces (red and blue lines respectively) and the nonadiabatic coupling vector (dashed line, multiplied by 30). %
}\label{fig:pop_LZ}
\end{figure}
To demonstrate this result numerically, we consider the Landau--Zener model for nonadiabatic transitions.\cite{Landau1932LZ,Zener1932LZ} This corresponds to initializing the bare electronic system in the upper adiabatic state for $t\rightarrow-\infty$ and then evolving the system under the pre-specified nuclear path, $q(t)=p_{\text{const}} t$, where $m=1$ and $p_{\text{const}}$ is the nuclear momentum, which is a constant during the entire dynamics. The diabatic Hamiltonian parameters [Eq.~(\ref{eq:ham_diab})] are given by 
\begin{subequations}
\begin{align}
\kappa(\bm{q})&=q , \\
\Delta(\bm{q})&=\Delta ,
\end{align}
\end{subequations}
from which the adiabatic Hamiltonian parameters, $V_{z}(\bm{q})$ and $d(\bm{q})$, needed for the mapping equations of motion [Eq.~(\ref{eq:mapping_eom})] can be generated using the formulas given in Appendix~\ref{sec:transf}. Additionally the long-time limit of the populations can be determined analytically for this model\cite{Zener1932LZ}
\begin{equation}
P^{\text{LZ}}_{+}(t\rightarrow\infty)=1-\eu{-\pi\Delta^{2}/p_{\text{const}}} ,
\end{equation}
which we would like to be able to reproduce with MASH\@. Figure~\ref{fig:pop_LZ} presents the resulting time-dependent adiabatic populations for $p_{\text{const}}/\Delta=2$, which show that the weighting factor $\mathcal{W}_{\text{PP}}(\bm{S})=2|S_{z}|$ is necessary for the binned adiabatic populations to reproduce the correct population transfer. We have also confirmed numerically that for this model, the two MASH expressions for this population--population correlation (green and blue lines) match their associated analytical expression given by Eqs.~(\ref{eq:pop_pop}) and (\ref{eq:pop_pop_weight}) respectively. 
These good results go beyond the Landau--Zener model.
For instance, in Sec.~\ref{sec:tully1}, we demonstrate that the accuracy of MASH is excellent for Tully's model I\@.
In particular, we note that this is only possible when correctly including the weighting factor, $\mathcal{W}_{\text{PP}}(\bm{S})$, and that if it were neglected the relative heights of the two peaks in the final distributions would be incorrect. %

Whilst we have only considered binning adiabatic populations in this paper, so as to ensure internal consistency with the nuclear force, the arguments presented in this section are equally applicable for other choices of the electronic basis functions. For example, binning the diabatic populations would result in an approach similar in spirit to that of SQC,\cite{Miller2016Faraday} especially if a mean-field expression for the force was additionally used. In this case, the following MASH expression for the diabatic population--population correlation would define the binning procedure: $|S^{\text{diab}}_{z}|h(S^{\text{diab}}_{z})h(S^{\text{diab}}_{z}(t))$.
\section{The Quantum--Classical Liouville Equation}\label{sec:QCLE}
In order to derive classical-trajectory approaches for simulating nonadiabatic dynamics and calculating real-time correlation functions, the first step is to replace the trace over the nuclear degrees of freedom by a phase-space average. This can be achieved in terms of the partial Wigner transforms of the initial and final observables, leading to the formally exact expression
\begin{equation}
C_{AB}(t)=\int\frac{\rd\bm{q}\rd\bm{p}}{(2\pi)^{f}}\,\tr\Big[\hat{A}(\bm{q},\bm{p})\hat{B}(\bm{q},\bm{p};t)\Big] ,
\end{equation}
where $\hat{A}(\bm{q},\bm{p})$ is the partial Wigner transform of the operator $\hat{A}$ over the nuclear degrees of freedom and $\tr[\cdot]$ is the quantum trace over the electronic degrees of freedom. Equations of motion for the time-evolved quantity $\hat{B}(\bm{q},\bm{p};t)$ can be obtained by taking the partial Wigner transform of both sides of the Quantum Liouville equation, $\rd\hat{B}/\rd t=\iu[\hat{H},\hat{B}]$. This leads to the so-called Moyal series and truncating this series at the first term leads to the quantum--classical Liouville equation (QCLE),\cite{Kapral1999,Bonella2010,Shi2004QCLE}
\begin{equation}
\label{eq:QCLE}
\begin{split}
&\frac{\rd\hat{B}(\bm{q},\bm{p};t)}{\rd t}\sim \iu\Big[\hat{V}_{\text{ad}}(\bm{q}),\hat{B}(\bm{q},\bm{p};t)\Big]+\sum_{j}\frac{p_{j}}{m}\frac{\partial \hat{B}(\bm{q},\bm{p};t)}{\partial q_{j}} \\ &+\sum_{j}\left(\bar{\mathcal{F}}_{j}(\bm{q})\frac{\partial \hat{B}(\bm{q},\bm{p};t)}{\partial p_{j}}+\frac{1}{2}\left[\hat{\mathcal{F}}_{j}(\bm{q}),\frac{\partial\hat{B}(\bm{q},\bm{p};t)}{\partial p_{j}}\right]_{+}\right) ,
\end{split}
\end{equation}
where $\hat{V}_{\text{ad}}(\bm{q})=V_{z}(\bm{q})\hat{\sigma}_{z}(\bm{q})+\sum_{j}\frac{d_{j}(\bm{q})p_{j}}{m}\hat{\sigma}_{y}(\bm{q})$ and $[\hat{A},\hat{B}]_{+}=\hat{A}\hat{B}+\hat{B}\hat{A}
$ is the anticommutator of electronic operators $\hat{A}$ and $\hat{B}$. Because these equations of motion correspond to the $\hbar\rightarrow0$ limit of the Moyal-series expansion of the Quantum Liouville equation, the QCLE describes the nonadiabatic dynamics associated with electrons coupled to classical nuclei. It can also be shown that, in general, the QCLE exactly reproduces quantum dynamics only in the short-time limit. However, it is exact for all times in the special case of systems consisting of electronic states linearly coupled to harmonic baths, such as the spin--boson model.\cite{MacKernan2002,Shi2004QCLE}

The QCLE cannot be solved using independent trajectories\cite{Kelly2012mapping,Kapral1999} and so the best possible outcome is to develop a classical-trajectory technique that constitutes the least severe approximation to it. Such approximations are commonly formulated by first representing the electronic degrees of freedom using mapping variables and the nuclear degrees of freedom by their classical phase-space variables. We would like such an approach to at least reproduce the QCLE to first-order in time, so that its accuracy could in principle be systematically converged to that of the QCLE by more frequent resampling of the associated electronic phase-space variables at intermediate times. To fulfill this criteria, a classical-trajectory approximation to the correlation function must satisfy $\dot{C}_{AB}(0)=\int\frac{\rd{\bm{q}}\rd{\bm{p}}}{(2\pi)^{f}}\,\tr\Big[\hat{A}(\bm{q},\bm{p})\dot{\hat{B}}(\bm{q},\bm{p})\Big]$ for arbitrary $\hat{A}$ and $\hat{B}$, where $\dot{\hat{B}}(\bm{q},\bm{p})$ is obtained from Eq.~(\ref{eq:QCLE}).  

For all fully-linearized spin-mapping approaches, including MASH, the correlation function is given by Eq.~(\ref{eq:corr_MASH}), although $\mathcal{W}_{AB}(\bm{S})=1$ for Ehrenfest, spin-LSC and FSSH\@. Additionally, the time-derivative of the mapped $\hat{B}(\bm{q},\bm{p})$ operator is
\begin{equation}
\label{eq:spin-deriv}
\begin{split}
&\dot{B}(\bm{q},\bm{p},\bm{S})=\frac{\partial B(\bm{q},\bm{p},\bm{S})}{\partial\bm{S}}\cdot\dot{\bm{S}}+\sum_{j}\frac{p_{j}}{m}\frac{\partial B(\bm{q},\bm{p},\bm{S})}{\partial q_{j}} \\
&+\sum_{j}\left(\bar{\mathcal{F}}_{j}(\bm{q})\frac{\partial B(\bm{q},\bm{p},\bm{S})}{\partial p_{j}}+\mathcal{F}_{j}(\bm{q},\bm{S})\frac{\partial B(\bm{q},\bm{p},\bm{S})}{\partial p_{j}}\right) .
\end{split}
\end{equation}
Note that $B(\bm{q}(t),\bm{p}(t),\bm{S}(t))=B(\bm{q},\bm{p},\bm{S};t)$ holds for any trajectory-based method, but not for the QCLE\@. Comparing the terms in Eqs.~(\ref{eq:QCLE}) and (\ref{eq:spin-deriv}), which are all first-order contributions to the respective approximate correlation function, we note that the first three terms on the right-hand side of Eq.~(\ref{eq:spin-deriv}) are an exact representation of the analogous QCLE terms. This results from constructing the mapping so that the bare electronic dynamics (as proved in Appendix \ref{sec:Rabi}) and the zero-time value of the correlation function are exactly reproduced. The final term on the right-hand side of Eq.~(\ref{eq:spin-deriv}), however, only exactly reproduces the QCLE term if the spin mapping additionally satisfies
\begin{equation}
\label{eq:force_crit_in}
\begin{split}
&\int\rd\bm{S}\,A(\bm{q},\bm{p},\bm{S})\mathcal{W}_{AB}(\bm{S})\mathcal{F}_{j}(\bm{q},\bm{S})\frac{\partial B(\bm{q},\bm{p},\bm{S})}{\partial p_{j}} \\
&\qquad\qquad=\frac{1}{2}\tr\Big[\hat{A}(\bm{q},\bm{p})\Big[\hat{\mathcal{F}}_{j}(\bm{q}),\frac{\partial\hat{B}(\bm{q},\bm{p})}{\partial p_{j}}\Big]_{+}\Big] ,
\end{split}
\end{equation}
for all possible operators $\hat{A}(\bm{q},\bm{p})$ and $\hat{B}(\bm{q},\bm{p})$ and for any nuclear mode, $j$. Because for two-level electronic subsystems the state-dependent nuclear force operator is given by Eq.~(\ref{eq:nuclear_force}), the only nonzero terms we need to consider are those for which the electronic contribution to one of either $\hat{A}(\bm{q},\bm{p})$ or $\partial\hat{B}(\bm{q},\bm{p})/\partial p_{j}$ is the identity and the other contribution is either given by $\hat{\sigma}_{x}(\bm{q})$ or $\hat{\sigma}_{z}(\bm{q})$. The terms that are identically zero need not be considered, because any operator symmetries are always reproduced by their spin-coordinate representations, guaranteeing by symmetry that these terms are also exactly zero within these approaches. Because of this, we replace Eq.~(\ref{eq:force_crit_in}) with the simpler requirement
\begin{equation}
\label{eq:force_crit}
\int\rd\bm{S}\,\mathcal{F}_{j}(\bm{q},\bm{S})\mathcal{W}_{\mathcal{I}\sigma_{k}}(\bm{S}) \sigma_{k}(\bm{S}) =\tr\Big[\hat{\mathcal{F}}_{j}(\bm{q})\hat{\sigma}_{k}(\bm{q})\Big] ,
\end{equation}
where $k\in\{x,z\}$ and $\mathcal{W}_{\mathcal{I}\sigma_{k}}(\bm{S})=\mathcal{W}_{\sigma_{k}\mathcal{I}}(\bm{S})$. This is automatically satisfied by spin-LSC as the Stratonovich--Weyl kernel guarantees that the mapping exactly reproduces the trace of products of any two electronic operators by the corresponding phase-space integral.\cite{spinmap} For MASH, using the expression for the state-dependent nuclear force [Eq.~(\ref{eq:force_MASH})] and the mapping representation of the correlation functions needed to exactly reproduce the dynamics in the classical path approximation [Eqs.~(\ref{eq:corr_MASH}) and (\ref{eq:MASH_weight})], it can be shown that Eq.~(\ref{eq:force_crit}) is also satisfied. Note that this result relies on the momentum-rescaling term, which we saw in Sec.~\ref{sec:momentum_rescaling} also contains momentum reversals at frustrated hops, and so this result would not hold if this term were neglected or modified arbitrarily. Therefore, both spin-LSC and MASH exactly reproduce the QCLE up to first-order in time and so both can rigorously make use of quantum-jump schemes to systematically improve the accuracy of their results to the QCLE for longer times.

We note that neither Ehrenfest nor FSSH satisfy Eq.~(\ref{eq:force_crit}) and so do not exactly reproduce the QCLE result up to first-order in time. This can be seen by considering the case $k=x$. To calculate this using either Ehrenfest or FSSH, the mapping variables would be sampled from both the north and south poles in accordance with Eq.~(\ref{eq:FSSH_sampling}), which would give zero on the left-hand side of Eq.~(\ref{eq:force_crit}) for $k=x$, because for this sampling $S_{x}=0$. Unlike spin-LSC and MASH, Ehrenfest and FSSH lack this rigorous connection to the QCLE and thus formally cannot make use of jump schemes to systematically improve the accuracy of their results to that of the QCLE.
\section{Microscopic Reversibility and Detailed Balance}\label{sec:detailed_balance}
Detailed balance and microscopic reversibility are important properties for dynamical approaches to obey, as they guarantee that the Boltzmann distribution is preserved for systems initialized in equilibrium.\cite{ellipsoid} In addition, they ensure that, for non-equilibrium initial conditions, the system correctly thermalizes in the long-time limit, assuming there is a large enough heat bath. Detailed balance is most easily defined in terms of equilibrium correlation functions of the form
\begin{equation}
\label{eq:equib_corr}
\mathcal{C}_{AB}(t)=\Tr\Big[\tfrac{1}{2}\big[\eu{-\beta\hat{H}},\hat{A}\big]_{+}\hat{B}(t)\Big] ,
\end{equation}
where $[\eu{-\beta\hat{H}},\hat{A}]_{+}$ symmetrizes the operators at time $t=0$. Note, we use calligraphic font for these equilibrium correlation functions to distinguish them from the nonequilibrium correlation functions [Eq.~(\ref{eq:corr_function})] used throughout the rest of the paper. If the dynamics conserves the Boltzmann operator, $\eu{-\beta\hat{H}}$, as is true in the quantum-mechanical case, it can be easily shown that
\begin{equation}
\label{eq:detailed_balance}
\mathcal{C}_{AB}(t)=\mathcal{C}_{BA}(-t) .
\end{equation}
We wish to know whether MASH also satisfies this condition and thereby obeys detailed balance. Note that this is not fully satisfied for Ehrenfest, FSSH, spin-LSC or SQC, although they may formally obey it in certain limits.  For instance, spin-LSC is known to correctly describe the long-time limit of correlation functions for systems that are not too asymmetric,\cite{ultrafast} SQC for weak system--bath coupling,\cite{Miller2015SQC} and FSSH for systems with weak nonadiabatic couplings.\cite{Schmidt2008}

For simplicity, we consider the case that both $\hat{A}$ and $\hat{B}$ are purely electronic operators. In MASH, the equilibrium correlation functions given by Eq.~(\ref{eq:equib_corr}) can be represented as follows
\begin{equation}
\begin{split}
\mathcal{C}^{\text{MASH}}_{AB}(t)=&\int\frac{\rd\bm{p}\,\rd\bm{q}}{(2\pi)^{f}}\int\rd\bm{S} \\
&\times\eu{-\beta E(\bm{p},\bm{q},\bm{S})}A(\bm{S})\mathcal{W}_{AB}(\bm{S})B(\bm{S}(t)) ,
\end{split}
\end{equation}
which is exact at time $t=0$ in the classical-nuclear limit. We consider the classical-nuclear limit as otherwise our approaches have the same problems with ZPE leakage as the classical Wigner method.\cite{Habershon2009water} Additionally the energy function, $E(\bm{p},\bm{q},\bm{S})$, is given by Eq.~(\ref{eq:energy_MASH}). In order to ascertain whether the detailed-balance condition given by Eq.~(\ref{eq:detailed_balance}) is satisfied by the MASH correlation functions, we use the fact that the MASH equations of motion are time-translationally invariant to give
\begin{equation}
\begin{split}
\mathcal{C}^{\text{MASH}}_{AB}(t)=&\int\frac{\rd\bm{p}(-t)\,\rd\bm{q}(-t)}{(2\pi)^{f}}\int\rd\bm{S}(-t) \\
&\times\eu{-\beta E(\bm{p},\bm{q},\bm{S})}B(\bm{S})\mathcal{W}_{AB}(\bm{S}(-t))A(\bm{S}(-t)) ,
\end{split}
\end{equation}
where we have additionally used the fact that $E(\bm{p},\bm{q},\bm{S})$ is conserved by the dynamics. Next we note that the MASH equations of motion are incompressible (i.e., $\pder{}{\bm{q}}\cdot\dot{\bm{q}}+\pder{}{\bm{p}}\cdot{\dot{\bm{p}}}+\pder{}{\bm{S}}\cdot\dot{\bm{S}}=0$),\cite{TuckermanBook} so that the dynamics preserves phase-space volume (i.e., $\rd\bm{p}(-t)\rd\bm{q}(-t)\rd\bm{S}(-t)=\rd\bm{p}\,\rd\bm{q}\,\rd\bm{S}$, which is a generalization of Liouville's theorem).
Therefore,
\begin{equation}
\label{eq:MASH_balance}
\begin{split}
\mathcal{C}^{\text{MASH}}_{AB}(t)=&\int\frac{\rd\bm{p}\,\rd\bm{q}}{(2\pi)^{f}}\int\rd\bm{S} \\
&\times\eu{-\beta E(\bm{p},\bm{q},\bm{S})}B(\bm{S})\mathcal{W}_{BA}(\bm{S}(-t))A(\bm{S}(-t)) ,
\end{split}
\end{equation}
where we have also used $\mathcal{W}_{AB}(\bm{S})=\mathcal{W}_{BA}(\bm{S})$ from Eq.~(\ref{eq:MASH_weight}). The right-hand side of Eq.~(\ref{eq:MASH_balance}) is almost but not identically equal to $\mathcal{C}^{\text{MASH}}_{BA}(-t)$, because the weighting factor, $\mathcal{W}_{AB}(\bm{S})$, is not in general conserved by the dynamics. From Eq.~(\ref{eq:MASH_weight}), we see that this is only a problem when calculating population--population correlation functions. However, because MASH does appear to accurately reproduce the long-time limit of correlation functions for condensed-phase systems in the classical-nuclear limit, as observed in Sec.~\ref{sec:spin-boson}, this suggests that detailed balance is at least approximately obeyed by the approach. This must mean that in practice the accuracy of MASH is not significantly affected by whether the weighting factor, $\mathcal{W}_{AB}(\bm{S})$, is applied at $t=0$ or at time $t$. For the regimes where MASH is exact (i.e., the electronic dynamics for a pre-defined nuclear path or the classical nuclear dynamics in the Born--Oppenheimer limit), we have already noted that using $\mathcal{W}_{AB}(\bm{S})$ or $\mathcal{W}_{AB}(\bm{S}(t))$ gives the same results (Appendix~\ref{sec:Rabi}) and so as expected microscopic reversibility is exactly obeyed in these cases. 

To test this hypothesis, the microscopic reversibility error can be defined [Eq.~(\ref{eq:error_function})] which corresponds to the difference in an observable obtained from weighting at these two different times. The MRE is calculated for many model systems in Sec.~\ref{sec:results} and is in general found to be small for the majority of regimes considered in this paper. For situations where the MRE does become large, the error can be reduced by resampling the spin-coordinates at intermediate times through the MASH jump procedure (Sec.~\ref{sec:jump}), thereby rigorously restoring microscopic reversibility and detailed balance within the dynamics.

%
\input{paper.bbl}

\end{document}

%% file: paper.bbl
%